\newcommand{\beq}{\begin{equation}} 
\newcommand{\eeq}{\end{equation}} 
\newcommand{\beqa}{\begin{eqnarray}} 
\newcommand{\eeqa}{\end{eqnarray}}

\def\simleq{\; \raise0.3ex\hbox{$<$\kern-0.75em
      \raise-1.1ex\hbox{$\sim$}}\; }
\def\simgeq{\; \raise0.3ex\hbox{$>$\kern-0.75em
      \raise-1.1ex\hbox{$\sim$}}\; }

\def\non{\nonumber }
\def\beq{\begin{equation} }
\def\eeq{\end{equation} }
\def\beqa{\begin{eqnarray}}
\def\eeqa{\end{eqnarray}}

\def\D{\Delta}
\def\a{\alpha }

\def\g{\gamma}

\def\med{\frac{1}{2}}

\def\eq{\!\! =\!\!}

\documentclass[preprint, tightenlines,eqsecnum,a4paper,amsmath,amssymb,superscriptaddress,nofootinbib]{revtex4}

\usepackage{bm}

\begin{document}

\title{Effective theory for the Goldstone field in  the \\ BCS-BEC crossover at $T=0$}

\author{Juan L.  Ma\~nes}
\affiliation
    {%
    Departamento de F\'\i sica de la Materia Condensada, 
    Universidad del Pa\'\i s Vasco, 
    Apartado 644, E-48080 Bilbao, Spain
    }%
\author{Manuel A. Valle}
\affiliation
    {%
    Departamento de F\'\i sica Te\'orica, 
    Universidad del Pa\'\i s Vasco, 
    Apartado 644, E-48080 Bilbao, Spain
    }%
\date{\today}

\begin{abstract} 
We perform a detailed study of the effective Lagrangian for the Goldstone mode of  a superfluid Fermi gas at zero temperature in the whole BCS-BEC crossover. 
By using a derivative expansion of the response functions, we derive the most general form of this Lagrangian at the next to leading order in the momentum expansion in terms of four coefficient functions.  This involves the elimination of all the higher order time derivatives by  careful use of the leading order field equations.   
In the infinite scattering length limit where conformal invariance is realized, we show that the effective Lagrangian must contain 
an unnoticed  invariant combination  of higher spatial gradients of the Goldstone mode, while  explicit couplings to spatial gradients of the trapping potential are absent. 
Across the whole crossover, we determine all the coefficient functions at the one-loop level, taking into account the dependence of the gap parameter on the chemical potential in the mean-field approximation. These results are analytically expressed in terms of elliptic integrals of the first and second kind.
We discuss the form of these coefficients in the extreme BCS and BEC regimes  and around the unitary limit, and compare with recent  work by other authors.

\end{abstract}


\maketitle


\section{Introduction}
\label{sec:intro}

The last few years have witnessed a renewed interest in the physics of the BCS-BEC crossover~\cite{Eagles,Leggett,Randeria}, 
partly motivated by the availability of  tunable interactions in the realm of interacting Fermi gases~\cite{Giorgini}.   
Recent experimental work~\cite{Regal,Zwierlein,Bartenstein,Kinast,Bourdel}  has shown evidence for condensation of 
fermionic atom pairs, suggesting  the formation of a fermionic superfluid. 
From the theoretical point of view, the qualitative description of the BCS-BEC crossover has been based on the mean-field theory of 
Leggett~\cite{Leggett} and  its extension to finite temperature by Nozi\`eres and Schmitt-Rink~\cite{Nozieres} and 
S\'a de Melo, Randeria and Engelbrecht~\cite{Melo}. 
In this description of the superconducting system, the effective action in terms of a  complex order parameter  
which couples to the pairing field  
plays a central role. 
Some recent developments~\cite{Pieri,Diehl1,Diehl2,Diener}  in the crossover problem beyond mean field theory have improved  
our understanding of the equilibrium state at a quantitative level.  In particular, 
Diener  {\it et al}.~\cite{Diener}, by computing  the complete   quadratic part of the effective action, have obtained 
the correction to the mean field result which arises from the integration of the Gaussian fluctuations, 
finding  excellent agreement with calculations based on quantum Monte Carlo techniques in the unitary limit 
where the scattering length  $a \rightarrow \infty$.


The effective action can also used, in principle,   to derive an effective Lagrangian which captures the low-energy behavior of the  system in terms
of the Goldstone mode phase of the order parameter. 
At zero temperature, one expects that an expansion of the effective Lagrangian in derivatives of the Goldstone field 
can be used in order to study the low-energy behavior of the system.
The leading order (LO) in this expansion was evaluated by Greiter {\it et al}.~\cite{Greiter} and by Aitchison {\it et al}.~\cite{Aitchison} some years ago, 
and since then,  
there have been various microscopic derivations~\cite{Palo,Benfatto,Dupuis,Liu,Rivers}  of effective models, some of which  ~\cite{Benfatto,Dupuis}  
have included more or less explicitly some derivative corrections.

In a recent work, Son and Wingate~\cite{Son} have systematically studied the form of the effective Lagrangian in the unitary limit
at the next-to-leading order (NLO) in the derivative expansion, when the effective theory is formulated 
in terms of the Goldstone mode coupled to external gauge and gravitational fields. At unitarity, it turns out that, 
besides general coordinate and gauge invariance, the theory exhibits  conformal invariance~\cite{Wise},  which puts constraints 
on the form of the NLO Lagrangian and restricts to two the number of independent NLO parameters. 

In this paper, we extend these studies. 
We evaluate the effective parameters of the NLO Lagrangian in the unitary limit  at the mean-field level,  
and also obtain the most general form of this Lagrangian away from this limit, where the symmetry under conformal transformations 
is not realized. By computing all the necessary functions at the one-loop level in terms of elliptic integrals, 
we have obtained the simplest approximation to the low energy effective theory  for the whole crossover region at zero temperature. 

We find  that, at unitarity,  the effective Lagrangian is specified by  two constants, 
but its actual form  differs from that given  in Ref.~\cite{Son}, and  includes
a new contribution $(\partial_i \partial_j \theta)^2$ of higher spatial derivatives of the Goldstone mode, 
while the NLO contribution of the external trapping potential proportional to $\nabla^2 V_{\text{ext}}$  is absent. We  show that these features are a necessary consequence of the conformal invariance of the NLO field equations. 
As an application, we derive the energy density functional  in the unitary limit, 
and compare it with the computation of Rupak and Sch\"affer~\cite{Rupak}, which is  
based on an epsilon expansion around $d=4-\varepsilon$ spatial dimensions and is, to our knowledge, the only one in the literature.  Altough the  coefficients computed by these two methods show  discrepancies of the order of $30\%$ in general,  
we find a surprisingly good agreement for the coefficient of the quantum pressure. 

For the whole crossover region, we obtain the NLO Lagrangian in terms of  four functions which are given in closed form in terms of elliptic integrals. The  BCS and BEC limits as well as the near unitarity limits of the NLO Lagrangian are worked out in detail. In the BEC limit, we recover the known features of the hydrodynamic description of superfluidity at zero temperature. 

The plan of this paper is as follows. In Section~\ref{sec:response}
we formulate the problem in the framework of linear response theory and 
 derive a linearized  equation for the Goldstone field in terms of derivatives of response functions.  
In Section~\ref{sec:higher} we show how to construct a Lagrangian, including second order time derivatives of $\theta$,  by considering all
the available Galilean invariants consistent with required general properties. 
Then we present a careful procedure of reduction, and  show how to use the LO field equation 
in order to eliminate  undesired higher order time derivatives without changing its perturbative contents.  We also
argue the need to compute two three-point functions in order to determine all the  coefficient functions in the effective Lagrangian. 
In Section~\ref{sec:para} we present   an analytical expression for the 
thermodynamical potential at the one-loop level, and  the analytical expressions of the NLO coefficients in the two 
(BCS and BEC) limits and near unitarity.
In Section~\ref{sec:functional} we compute the energy density functional in the unitary limit, 
and  compare our result with other approaches. 
Section~\ref{sec:end} gives our conclusions. 
Details of the calculations as well as additional material on the  invariance properties under conformal transformations in the unitary limit are 
given in a series of four appendices.


\section{Derivative expansion of the response functions at $T=0$}
\label{sec:response}

The system is conveniently  described by the BCS Lagrangian density in terms of a Nambu spinor field and a 
complex field which decouples the short-range interaction in the Cooper channel
\begin{equation}
\label{eq:lag}
\mathcal{L} = \Psi^\dagger \left[ i \partial_t + \tau^3 \frac{1}{2 m} \nabla^2 
+ \frac{1}{2}(\tau^1 + i \tau^2) \Delta + \frac{1}{2}(\tau^1 - i \tau^2) \Delta^\ast  \right]  \Psi - 
\frac{1}{g_\Lambda}  \Delta^\ast \Delta ,
\end{equation}
where $\Psi^\dagger = (\psi_\uparrow^\dagger, \psi_\downarrow)$ and  $\tau^i$ is  the corresponding Pauli matrix. 
The bare coupling parameter $g_\Lambda$ depends on the details of the short range interaction and on a regulator that truncates some loop integrals at some very large scale $\Lambda$.  As usual in the framework of effective field theories, the result of adding the appropriate loop corrections to this bare coupling  will be matched  with the measured low energy scattering properties encoded in the $s$-wave scattering length  $a$. 
Other  physical parameters such as the $s$-wave effective range or 
the $p$-wave scattering length are related to operators with more derivatives in the effective Lagrangian, and hence,  they are generically subleading in the expansion in powers of $R \partial$,  where $R$ is some length setting the size of the interaction region. 
The Lagrangian is invariant under the $U(1)$ symmetry of phase independent spacetime-independent transformations 
$\Psi \rightarrow  e^{i \tau^3 \theta} \Psi$, $\Delta \rightarrow  e^{i 2  \theta} \Delta$, which is spontaneously broken down to $H=Z_2$ below a critical temperature. 
In order to compute the effective action for the resulting gapless collective mode it is convenient~\cite{Weinberg}  to express the fields 
as a $U(1)$ transformation acting on fields $\widetilde{\Psi}$ and $\sigma$ which do not contain the Goldstone mode  
\begin{equation}
\Psi(x) = e^{i \tau^3 \theta(x)} \widetilde{\Psi}(x), \quad \Delta(x) = e^{i 2  \theta(x)} (\Delta_0 + \sigma(x)) , 
\end{equation} 
where the constant amplitude $\Delta_0$ and  its fluctuation $\sigma(x)$ are real numbers. 
Since we shall hereafter use the  fermion fields  $\widetilde{\Psi}$, the tildes will be dropped to simplify the notation. 
With this choice, the Lagrangian involving the couplings between the fermion field and the other fields becomes, after an integration by parts,
\begin{equation}
\label{eq:LI}
\mathcal{L}_I =- \Psi^\dagger \tau^3 \Psi \left(\partial_t \theta+ 
  \frac{1}{2 m}  (\bm{\nabla} \theta)^2 +V_\text{ext}\right )   
 -  \frac{1}{2m i}(\Psi^\dagger \bm{\nabla} \Psi-\bm{\nabla}\Psi^\dagger \Psi)\cdot \nabla \theta 
  + \Psi^\dagger \tau^1 \Psi \sigma , 
\end{equation}
where $V_\text{ext}$ is an  external potential. 
The quadratic part $\mathcal{L}_2$  of the Lagrangian including the chemical potential coupled to conserved particle density  $\Psi^\dagger \tau^3 \Psi$ is given by 
Eq.~(\ref{eq:lag})  with the replacements  $\nabla^2/(2m) \rightarrow \nabla^2/(2m) +\mu$,  $\Delta  \rightarrow \Delta_0$ and $\Delta^\ast \Delta \rightarrow \sigma^2$. 

Under  an infinitesimal Galilean transformation 
\begin{equation}
     \bm{x} \rightarrow \bm{x}' = \bm{x} + \bm{v} t, \qquad t \rightarrow t' = t ,
\end{equation}
the Goldstone field changes inhomogeneously as 
\begin{equation}
\delta \theta(t,\bm{x}) = -\bm{v} t \cdot \bm{\nabla} \theta+ m \bm{v} \cdot \bm{x} . 
\end{equation}
Once the auxiliary field $\sigma$ has been eliminated, 
the invariance of the effective action for $\theta$ requires a dependence through the scalar quantity
\begin{equation}
X\equiv\mu - \partial_t \theta - \frac{1}{2 m}  (\bm{\nabla} \theta)^2 - V_\text{ext}, \quad 
\delta X = - \bm{v} t \cdot \bm{\nabla} X , 
\end{equation}
and some appropriate derivatives of $X$ and $\bm{\nabla}\theta$. 
According to the power counting scheme of Ref.~\cite{Son}\footnote{A quantity containing $N[\partial_t]$ time derivatives, 
$N[\partial_i]$ spatial derivatives and  $N[\theta]$ powers of the Goldstone field is counted as $O(p^N)$, where $N=N[\partial_t]+ N[\partial_i]-N[\theta]$.}, 
these quantities are $O(p^0)$. This can also be seen  by noting that, at equilibrium, 
the gradient of the phase $\bm{v}_s =  m^{-1} \bm{\nabla} \theta$ is  the constant superfluid velocity, 
and $\partial_t \theta$ goes with the chemical potential.  
To next-to-leading order, we have the following $O(p)$ galilean invariant derivatives 
\begin{equation}
\label{eq:one}
\nabla^2 \theta, \quad  \left[\partial_t + \frac{\bm{\nabla} \theta}{m} \cdot \bm{\nabla} \right] X, 
\quad \bm{\nabla} X, \quad \partial_i \partial_j \theta, 
\end{equation}
where we have only written derivatives of the quantities $X$ and $\bm{\nabla}\theta$, which are coupled to the particle density and the current. The signature under time reversal ($t\to -t$, $\theta\to -\theta$) of the two  scalar terms (and the tensorial term) is $-1$, and hence there is no contribution of $O(p)$ to the effective Lagrangian.   In the next Section we will list  all possible scalar terms of $O(p^2)$ potentially contributing to the effective Lagrangian.

A possible way to derive the effective Lagrangian for the Goldstone mode is to compute the appropriate linear response at low frequency and momentum for an external perturbation given by a Hamiltonian 
\begin{eqnarray}
\label{eq:ham}
 H^\text{ext}(t) &=&  \int d^3 x \left[  \Psi^\dagger \tau^3 \Psi ( \partial_t \theta + V_\text{ext})   + 
     \frac{1}{2m i}(\Psi^\dagger \bm{\nabla} \Psi-\bm{\nabla}\Psi^\dagger \Psi)\cdot \bm{\nabla} \theta  - 
     \Psi^\dagger \tau^1 \Psi \sigma \right]  \nonumber \\
      &=&  \int d^3 x \left[ n ( \partial_t \theta + V_\text{ext})  + \bm{j}^\text{p} \cdot \bm{\nabla} \theta -   \Psi^\dagger \tau^1 \Psi \sigma \right], 
\end{eqnarray}
where $\bm{j}^\text{p}$ is the current density operator in the absence of $\bm{\nabla} \theta$  and the total current operator is 
$\bm{j} = \bm{j}^\text{p} +\Psi^\dagger \tau^3 \Psi m^{-1} \bm{\nabla} \theta$.  
The induced changes to be computed are $\delta \langle n(Q) \rangle$, $\delta \langle \bm{j}(Q) \rangle$ and 
the  change in the expectation value of the pairing field $\delta \langle \Psi^\dagger \tau^1 \Psi \rangle $
in terms 
of  $\theta(Q)$ and  $\sigma(Q)$, where $Q = (\bm{q}, \omega)$.  
By combining the conservation of the particle number with the gap equation for $\delta \langle \Psi^\dagger \tau^1 \Psi \rangle $, 
one can obtain  a single linear equation of motion for  $\theta$ and, consequently, a quadratic Lagrangian to be consistently matched with the form of the Galilean invariants listed above. 

In what follows, we use $\chi_{AB}(Q)$ for the Fourier transform of the retarded response function $\chi_{AB}(X, X') = -i \langle \left[A(X), B(X')\right] \rangle \theta(t-t')$. 
For the Hamiltonian given in Eq.~(\ref{eq:ham}), the required set of linear response equations is
\begin{eqnarray}
\delta \langle n(Q)\rangle &=& \chi_{nn}(Q) \left(-i \omega \theta(Q) + V_\text{ext}(Q) \right) + \chi_ {n j}^{ \;\; k}(Q) i q^k \theta(Q)  -
 \chi_{n1}(Q) \sigma(Q), \\ 
\delta \langle j^k(Q)\rangle &=& \chi_{j n}^{k}(Q)  \left(-i \omega \theta(Q) + V_\text{ext}(Q) \right) + \chi_ {j j} ^{k l}(Q) i q^l \theta(Q) -
 \chi_{j 1}^{k} (Q) \sigma(Q) , \\
\delta \langle \Psi^\dagger \tau^1 \Psi \rangle 
 &=& \chi_{1 n}(Q)  \left(-i \omega \theta(Q) + V_\text{ext}(Q) \right) + \chi_ {1j}^{\; \; k}(Q) i q^k \theta(Q)  -
 \chi_{11}(Q) \sigma(Q), 
\end{eqnarray} 
where there is a summation over  upper repeated indices. 
It is useful to collect some symmetry properties that follow from the behavior under time reversal and parity: 
\begin{eqnarray}
 \chi_{j n}^{k}(Q)&=& \chi_ {n j}^{ \;\; k}(Q)=  \chi_ {j n}(Q) q^k, \\ 
 \chi_{j 1}^{k} (Q) &=& \chi_ {1 j}^{ \;\; k}(Q) =  \chi_ {j1}(Q) q^k, \\ 
 \chi_{n1}(Q)&=& \chi_ {1n}(Q).  
\end{eqnarray}

When the operators $A$ and $B$ have the same (opposite) signature under time reversal, 
$\text {Im} \chi_ {A B}(Q)$ is  odd (even) in $\omega$. This implies that 
$\text {Re } \chi_ {j n}(Q)$, $\text {Re }\chi_ {j1}(Q)$ are odd in $\omega$,  while  
 $\text {Re} \chi_ {n1}(Q)$, $\text {Re} \chi_ {nn}(Q)$ and $\text {Re} \chi_ {1 1 }(Q)$ are  even functions of $\omega$. 
 The current-current response function can be written as 
 \begin{equation}
 \chi_ {j j} ^{k l}(Q)  = \chi_\text{L}(Q) \hat{q}^k  \hat{q}^l + \chi_\text{T}(Q) \left(\delta^{k l} - \hat{q}^k  \hat{q}^l\right), 
 \end{equation}
 where $\text {Re} \chi_\text{L,T}(Q)$ are even in $\omega$. 
At $\bm{q}=0$,  they must obey the usual sum rules giving the static responses at superfluidity
\begin{eqnarray}
\chi_{\text{L}}(0) &=& \lim_{\bm{q} \rightarrow 0}
    \int_{-\infty}^\infty \frac{d\omega}{\pi} \frac{\text{Im} \chi_\text{L}(\bm{q},\omega)}{\omega} = \frac{\langle n \rangle}{m} , \\ 
\label{eq:trule}
\chi_{\text{T}}(0) &=& \lim_{\bm{q} \rightarrow 0}
\int_{-\infty}^\infty \frac{d\omega}{\pi} \frac{\text{Im} \chi_\text{T}(\bm{q},\omega)}{\omega} = 
  \frac{\langle n \rangle - n_s}{m}, 
\end{eqnarray}
where $n_s$ is the superfluid particle density. 

The results of this Section do not depend on the specific approximation used to compute the response functions, but in order to gain some physical insight on them, we write the kind of integrals to be computed at the one-loop level 
\begin{equation}
\label{eq:int}
\left. \int  \frac{d \nu_1 d^3 k}{(2\pi)^4} 
\left\{
\begin{array}{c}
1  \\
(2\bm{k}+\bm{q})^k \\
(2\bm{k}+\bm{q})^k (2\bm{k}+\bm{q})^l
\end{array}
\right\} 
 \text{tr}\left[\tau^\mu \mathcal{G}(\bm{k}+\bm{q}, i \nu_1+ z) \tau^\nu
    \mathcal{G}(\bm{k}, i \nu_1)\right]  \right|_{z = \omega + i \varepsilon}. 
\end{equation}
Here  $\tau^0 $ is the identity matrix and  $\mathcal{G}$ is the Nambu-Gorkov Green's function 
\begin{equation}
\mathcal{G}(\bm{k}, z) = \frac{z + \tau^3 \xi_{\bm{k}}  - \tau^1 \Delta_0}{z^2 - E_{\bm{k}}^2}, 
\end{equation} 
with  $\xi_{\bm{k}} =  \epsilon_{\bm{k}}-\mu =\bm{k}^2/2m - \mu$ and $E_{\bm{k}}= \sqrt{\xi_{\bm{k}}^2 + \Delta_0^2}$. 
Note that, together with the contribution coming from $[\bm{j}^\text{p},\bm{j}^\text{p}]$, the current-current response function $\chi_ {j j} ^{k l}(Q)$ includes the longitudinal piece $\langle n\rangle m^{-1}  \hat{q}^k \hat{q}^l$, where $\langle n \rangle$ is the total particle density. 
The integration over the imaginary frequency produces the denominators $ \omega + i \varepsilon \pm (E_{\bm{k}} + E_{\bm{k}+\bm{q}})$, 
which for  small frequency $|\omega| < 2 \Delta_0$ do not contribute to the imaginary part of the response function. 
In addition, the expansions about $\omega=0$ and $\bm{q}=0$ are well behaved,  in marked contrast to the expansions of the denominators $ \omega + i \varepsilon \pm (E_{\bm{k}} - E_{\bm{k}+\bm{q}})$  which arise from  Landau damping at $T \neq 0$. 
Thus, at $T=0$ the needed response functions are regular real functions near $Q=0$.  
After some calculation, the one-loop expressions for $\chi_{AB}(\bm{q}=0, \omega=0)$ become 
\begin{eqnarray}
\label{eq:nn}
\chi_{n n}(0) &=&-\int \frac{d^3 k}{(2 \pi)^3} \frac{\Delta_0^2}{E_{\bm{k}}^3}, \\ 
\label{eq:n1}
\chi_{n 1}(0) &=&-\int \frac{d^3 k}{(2 \pi)^3} \frac{ \Delta_0 \xi_{\bm{k}} }{E_{\bm{k}}^3}, \\ 
\label{eq:11}
\chi_{1 1}^{(\Lambda)}(0) &=&-\int^\Lambda \frac{d^3 k}{(2 \pi)^3} 
 \frac{ \xi_{\bm{k}}^2 }{E_{\bm{k}}^3}, \\ 
\label{eq:jj1}
\chi_ {j j} ^{k l}(0) & = & \frac{\langle n\rangle}{m}  \hat{q}^k \hat{q}^l , 
\end{eqnarray}
and we see that static  current transverse response $\chi_\text{T}(0)$ vanishes which, according to Eq.~(\ref{eq:trule}), 
shows that the entire system is superfluid,  $\langle n \rangle = n_s$, at $T=0$.

The  linear ultraviolet divergence in $\chi_{1 1} $ has been regulated by the cut-off $\Lambda$. 
As it is well known, the renormalization of this divergence is performed 
by the substitution of the bare coupling constant $g_\Lambda$  in terms of the $s$-wave scattering length $a$.   Putting together 
the leading order interaction of the form $g_\Lambda^{-1}$ and the first correction in the vacuum, one obtains the 
measured coupling constant  $g \equiv -4 \pi a m^{-1}$, 
\begin{equation}
  \frac{1}{g}= \frac{1}{g_\Lambda} - \int^\Lambda \frac{d^3 k}{(2 \pi)^3} \frac{1}{2 \epsilon_{\bm{k}}}, 
\end{equation} 
and this relation allows the determination of $g_\Lambda$ in terms of $a$. As all the calculations will be $\Lambda$ independent,  
it is convenient to define the `renormalized'  $\chi_{1 1} $ as
\begin{equation}
\chi_{1 1}(0) \equiv - \int \frac{d^3 k}{(2 \pi)^3} \left( \frac{ \xi_{\bm{k}}^2 }{E_{\bm{k}}^3} -\frac{1}{\epsilon_{\bm{k}}} \right) = 
\chi_{1 1}^{(\Lambda)}(0)  + \frac{2}{g_\Lambda} + \frac{m}{2 \pi a} .
\end{equation}

As mentioned above, at $T=0$  all response functions are real in the region of small $Q$  and their lowest-order derivatives at $Q=0$ must determine the form of the effective Lagrangian at next-to-leading order in the derivative expansion. We next show how to compute this Lagrangian.  
From the gap equation 
\begin{equation}
\delta \langle \Psi^\dagger \tau^1 \Psi \rangle =  \frac{2}{g_\Lambda} \sigma(Q) ,
\end{equation}
and the equation for the response $\delta \langle \Psi^\dagger \tau^1 \Psi \rangle$, 
we  can write the change $\sigma(Q)$ in the local amplitude  in terms of  $\theta(Q)$ and $V_\text{ext}(Q)$.  
If we replace the result for $\sigma(Q)$ in the continuity equation 
\begin{equation}
-\omega \delta \langle n(Q)\rangle +  q^k \delta \langle j^k(Q)\rangle = 0, 
\end{equation}
and expand in powers of $\omega$ and $q$,  we find the following equation for $\theta(Q)$  
\begin{eqnarray}
\label{eq:cont}
0 &=& \left(b_1 \omega^2 + b_2 q^2 + b_3 \omega^2 q^2 + b_4 \omega^4 + b_5 q^4 + \ldots\right) \theta(Q) \nonumber \\
        &&+ \left(b_1  + b_6  q^2 + b_4  \omega^2 + \ldots \right)  i \omega V_\text{ext}(Q) , 
\end{eqnarray}
where 
\begin{subequations}\label{eq:bes}
\begin{eqnarray}
\label{eq:b1}
b_1 &=&   -\chi_{n n}(0) + \frac{g \chi_{n 1}(0)^2}{2 + g \chi_{1 1}(0)} , \\ 
\label{eq:b2}
b_2 &=&-\chi_\text{L}(0)= -\frac{\langle n \rangle}{m},  \\ 
b_3 &=& 2 \frac{\partial \chi_{j n}}{\partial \omega} - \frac{\partial \chi_{nn}}{\partial q^2}-\frac{\partial \chi_\text{L}}{\partial \omega^2}  -
  \frac{2 g \chi_{n 1}}{2 + g \chi_{1 1}} \left( \frac{\partial \chi_{j 1}}{\partial \omega} - \frac{\partial \chi_{n1}}{\partial q^2}\right) -
  \frac{g^2 \chi_{n 1}^2}{(2 + g \chi_{1 1})^2} \frac{\partial \chi_{11}}{\partial q^2} , \\ 
b_4 &=&  -\frac{\partial \chi_{nn}}{\partial \omega^2}  +
    \frac{2 g \chi_{n 1}}{2 + g \chi_{1 1}} \frac{\partial \chi_{n1}}{\partial \omega^2}  
    - \frac{g^2 \chi_{n 1}^2}{(2 + g \chi_{1 1})^2}  \frac{\partial \chi_{11}}{\partial \omega^2}, \\
b_5 &=& -\frac{\partial \chi_\text{L}}{\partial q^2} , \\ 
b_6 &=& \frac{\partial \chi_{j n}}{\partial \omega} - \frac{\partial \chi_{nn}}{\partial q^2}-
  \frac{2 g \chi_{n 1}}{2 + g \chi_{1 1}} \left(\frac{1}{2} \frac{\partial \chi_{j 1}}{\partial \omega} - \frac{\partial \chi_{n1}}{\partial q^2}\right) -
  \frac{g^2 \chi_{n 1}^2}{(2 + g \chi_{1 1})^2} \frac{\partial \chi_{11}}{\partial q^2} . 
\end{eqnarray}
\end{subequations}
All  these parameters are computable from the appropriate two-point retarded functions and their  derivatives. 
As the gap equation in terms of the  thermodynamic potential $\Omega(\mu, \Delta_0)$ at $T=0$  
\begin{equation}
\label{eq:gap}
\frac{\partial\Omega(\mu, \Delta_0)}{\partial \Delta_0} = 0 
\end{equation} 
implicitly determines the function $\Delta_0(\mu)$, 
the coefficients $b_i(\mu)$  depend on the chemical potential and,  parametrically,  
on the scattering length $a$. 


\section{The higher-order and the reduced effective Lagrangians}
\label{sec:higher}

\subsection{The higher-order effective Lagrangian}
We now  write an effective Lagrangian for the Goldstone mode   
which leads to the above equation of motion  for $\theta$. This Lagrangian contains higher-order time derivatives, which will be dealt with in the second part of this Section. Once the auxiliary field $\sigma$ has been eliminated through the use of the gap equation, the 
NLO Lagrangian is a linear combination of all the independent scalar operators of  $O(p^2)$ 
constructed from   derivatives of $X$ and $\bm{\nabla}\theta$, with coefficients which are functions of $X$,  the only scalar  of  $O(p^0)$. 
Note that, as mentioned above,    $O(p)$ scalar terms such as $r_1(X)\nabla^2 \theta$ s and $r_2(X) X_t $ are excluded by invariance under time reversal. 

There are only five independent NLO terms, 
$(\bm{\nabla} X )^2, (\nabla^2 \theta)^2, (\partial_i \partial_j \theta)^2, X_t^2$,  and $X_t  \nabla^2 \theta$, where 
\beq
X_t\equiv \partial_t  X+ m^{-1} \bm{\nabla} \theta \cdot \bm{\nabla} X\,.
\eeq
Other candidate terms can be shown  to be dependent on these five. For instante,  
we have excluded a term  $h(X) \nabla^2 X$, which becomes proportional to $(\bm{\nabla} X )^2$ after an  integration by parts. 
Another potential contribution of  $O(p^2)$, $(\partial_t + m^{-1} \bm{\nabla} \theta\cdot\bm{\nabla})\nabla^2\theta$, which by  
\[
(\partial_t + m^{-1} \bm{\nabla} \theta\cdot\bm{\nabla})\nabla^2 \theta = -\nabla^2 X-\nabla^2 V_\text{ext} - m^{-1} (\partial_i \partial_j \theta)^2
\] 
is equivalent to $\nabla^2 V_\text{ext}$, 
must also be excluded, since the identity 
\begin{eqnarray}
\label{eq:identity}
&& G'(X) X_t  \nabla^2 \theta + G'(X) (\bm{\nabla} X )^2 - G(X) \nabla^2 V_\text{ext} - 
\frac{1}{m} G(X) \left[(\partial_i \partial_j \theta)^2 - (\nabla^2 \theta )^2 \right]  = \nonumber \\ 
&& \quad \partial_t \left(G(X) \nabla^2 \theta \right) + \bm{\nabla} \cdot \left(\frac{ \bm{\nabla} \theta}{m}\, G(X) \nabla^2\theta  +
 G(X) \bm{\nabla} X \right)   
\end{eqnarray}  
shows that this term is in fact redundant.  
Thus,  the most general  Lagrangian up to  next-to-leading order in the derivative expansion is given in terms of 
six coefficient functions  
\begin{equation}
\label{eq:lag6}
\mathcal{L} = P(X)+g_1(X) (\bm{\nabla} X )^2 + g_2(X) (\nabla^2 \theta )^2
          + g_3(X) (\partial_i \partial_j \theta)^2+g_4(X) X_t ^2 +  g_5(X) X_t  \nabla^2 \theta  ,  
\end{equation}
to be evaluated  from the low-energy behavior of the response functions. 

The linearized field equation  following from this Lagrangian  has the form
\begin{eqnarray}
\label{eq:lin}
0 &=& -P''(\mu) \left(\frac{\partial^2 \theta}{\partial t^2} + \frac{\partial V_\text{ext}}{\partial t} \right) +
\frac{P'(\mu)}{m} \nabla^2 \theta  + 
  2\left[  g_1(\mu)- g_5(\mu)\right ] \frac{\partial^2 (\nabla^2 \theta)}{\partial t^2} \nonumber \\ 
  &&+2\left[  g_2(\mu) + g_3(\mu)\right ]  \nabla^4 \theta  
    + 
     \left[ 2 g_1(\mu) -g_5(\mu) \right]  \frac{\partial (\nabla^2 V_\text{ext})}{\partial t} 
     + 2 g_4(\mu)\left(\frac{\partial^4 \theta}{\partial t^4} + 
                       \frac{\partial^3 V_\text{ext}}{\partial t^3} \right).   
\end{eqnarray} 
Comparison with Eq.~(\ref{eq:cont}) yields  the  relations 
\begin{subequations}
\label{eq:gi}
\begin{eqnarray}
g_1 &=& -\frac{b_3}{2} + b_6 , \\ 
g_2 +g_3&=& \frac{b_5}{2} , \\    
g_4 &=&\frac{b_4}{2},   \\ 
g_5 &=&-b_3+b_6,
\end{eqnarray}
\end{subequations}
and 
\begin{subequations}
\begin{eqnarray}
 P''(\mu) &=& b_1 , \\ 
 P'(\mu) &=& -m b_2 = \langle n \rangle . 
\end{eqnarray}
\end{subequations}

These expressions match the coefficients of the effective Lagrangian with the response functions. Note that from  the response functions we can only 
determine the sum \hbox{$g_2(X)+g_3(X)$}, but not the individual functions. This is not surprising  
since, in the quadratic approximation to the  Lagrangian, the terms $ (\nabla^2 \theta)^2$ and $(\partial_i \partial_j \theta)^2$ are not independent, but differ by a total derivative.    
Hence, in order to evaluate  $g_2$ and $g_3$ separately, one must  resort to the computation of   three-point functions. 
Details of this computation at the one-loop level are given in  Appendix~\ref{sec:3point},  where we find the remarkably simple result \hbox{$g_3(X)/g_2(X) = 2$}. This implies:
\beq\label{eq:g2g3}
g_2(X)={1\over 6} b_5\;\; , \; \; g_3(X)={1\over 3} b_5 .
\eeq

Regarding the expressions for the leading order coefficients $b_1(\mu)$ and $b_2(\mu)$,  
it is possible to check their consistency without actually computing  them. 
As the authors of Refs.~\cite{Greiter,Son} have shown, the  leading order effective Lagrangian at $T=0$  is 
a function of the invariant $X$ which precisely coincides with the pressure $P(\mu)$ as a function of the chemical potential. 
Obviously, the expression for $b_2$ in~(\ref{eq:b2}) satisfies this  requirement.  
To see the agreement of $b_1$ in~(\ref{eq:b1}) with $P''(\mu) $, we note that the  pressure is given by 
$P(\mu) =-V^{-1}\Omega(\mu, \Delta_0(\mu))$,  
where $\Delta_0(\mu)$ satisfies the gap equation~(\ref{eq:gap}). 
The thermodynamic potential can be written as the sum of a `tree'  level renormalized piece and the full quantum contribution $\Phi$ from the perturbative expansion
\begin{equation}
\frac{\Omega(\mu, \Delta_0)}{V} =  \frac{\Delta_0^2}{g} + \Phi(\mu, \Delta_0). 
\end{equation} 
By using  
$\Delta_0'(\mu) = -\left(\partial^2 \Omega/ \partial \Delta_0^2 \right)^{-1}\partial^2 \Omega/\partial \mu \partial \Delta_0$, a direct evaluation  of $P''(\mu)$ yields
\begin{equation}
P''(\mu) = \left. -V^{-1}  \frac{\partial^2 \Omega}{\partial \mu^2} + V^{-1} \left (\frac{\partial^2 \Omega}{\partial \mu \partial \Delta_0} \right)^2 
 \left(\frac{\partial^2 \Omega}{\partial \Delta_0^2}  \right)^{-1} \right|_{\Delta_0 = \Delta_0(\mu)},  
\end{equation}
which agrees with the expression for $b_1$ when one identifies properly the static response functions  with the susceptibilities
\begin{eqnarray}
\chi_{n n}(0) &=& V^{-1}  \frac{\partial^2 \Omega}{\partial \mu^2} , \\ 
\chi_{n 1}(0) &=& V^{-1}  \frac{\partial^2 \Omega}{\partial \mu \partial \Delta_0} , \\ 
\chi_{1 1} (0)&=& V^{-1}  \frac{\partial^2 \Omega}{\partial \Delta_0^2} - \frac{2}{g} .
\end{eqnarray}
Note that the condition of thermodynamic stability implies that  $b_1$ must be positive. 
This analysis shows the important role played by  the response functions of  the pairing field  $\Psi^\dagger \tau^1 \Psi$ 
in the construction of an effective Lagrangian satisfying  known  general properties. Some recent work~\cite{Liu} has  overlooked this point. 

\subsection{Reduction of the higher-order Lagrangian}

To proceed further, we must find a ``reduced effective Lagrangian" without higher-order time derivatives, but  perturbatively equivalent  to Eq.~(\ref{eq:lag6}). The LO field equations can be used, in principle, to eliminate terms with higher order time-derivatives in favour of terms with spatial derivatives but, in doing so, one must be very careful that  the perturbative content of the original Lagrangian is preserved. In this regard, it is important to note that the LO equations 
\begin{equation}
\label{eq:zero}
X_t  +\frac{P'(X)}{m P''(X)} \nabla^2 \theta = O(p^3) 
\end{equation}
are satisfied only up to terms of  $O(p^3)$ coming from the NLO Lagrangian.

The term proportional to    $X_t^2 $,  which gives rise to  fourth-order time derivatives of $\theta$ in the field equation, 
can be eliminated by adding and subtracting $g_4(X)$ times the square of the LO field equations. 
This yields
\begin{eqnarray}\label{eq:quad}
\mathcal{L} &=&P(X)+ g_1(X) (\bm{\nabla} X )^2 + g_3(X) (\partial_i \partial_j \theta)^2 \nonumber \\
           && + \left[g_2(X) -
             g_4 (X)\left( \frac{P'(X)}{m P''(X)} \right)^2 \right] (\nabla^2 \theta )^2 
            \nonumber  \\ 
           &&  +\left[g_5(X)  - 2 g_4(X) \frac{P'(X)}{m P''(X)} \right] X_t  \nabla^2 \theta 
           +g_4(X) \left[ X_t  +\frac{P'(X)}{m P''(X)} \nabla^2 \theta \right]^2 . 
\end{eqnarray}
Now, the term proportional to the square of the LO  field equation, when evaluated on a  perturbative solution,  is  of $O(p^6)$, and thus highly suppressed.  Furthermore, one can easily check that  its contribution to the field equation is of   $O(p^5)$. It can thus be safely dropped and we are left with the following reduced Lagrangian   
\begin{equation}\label{eq:int}
\widetilde{\mathcal{L}} =P(X) +g_1(X) (\bm{\nabla} X )^2 +  
 \widetilde{g}_2(X)(\nabla^2 \theta )^2 + g_3(X) (\partial_i \partial_j \theta)^2+\widetilde{g}_5(X) X_t  \nabla^2\theta .
\end{equation}

We still have to get rid of the second-order time derivative in the last term of 
$\widetilde{\mathcal{L}}_\text{NLO}$. 
But in this case we can not use the leading field equation 
\begin{equation}
\label{eq:arg}
\widetilde{g}_5(X) X_t  \nabla^2 \theta \equiv \widetilde{g}_5(X)
\left( X_t  +\frac{P'(X)}{m P''(X)} \nabla^2 \theta \right) \nabla^2 \theta - 
\widetilde{g}_5(X)\frac{P'(X)}{m P''(X)}(\nabla^2 \theta)^2  
\end{equation}
to this end since, even though the numerical correction to the classsical effective action\footnote{By this we mean the effective action $\int d^3x dt\widetilde{\mathcal{L}}$ evaluated on a classical solution.} introduced  
by dropping the first term in the RHS of Eq.~(\ref{eq:arg}) would be of  $O(p^4)$, and thus acceptable, 
the corresponding correction to the field equation would be  $O(p^3)$, which is of the same order as the NLO contributions to the field equations.  
This can be checked explicitly by computing the linearized correction to the field equation  coming from the first term of the RHS of Eq.~(\ref{eq:arg}) \begin{equation}
-2 \widetilde{g}_5(\mu) \nabla^2 \left( \frac{\partial^2 \theta}{\partial t^2}  +  
\frac{\partial V_\text{ext}}{\partial t} - \frac{P'(\mu)}{m P''(\mu)}  \nabla^2 \theta \right) + 
\widetilde{g}_5(\mu)  \frac{\partial (\nabla^2 V_\text{ext})}{\partial t}.
\end{equation}
While, by  Eq.~(\ref{eq:zero}), the first term is $O(p^5)$  on a classical  solution and can be dropped, 
the second term yields a contribution of  $O(p^3)$ which produces an unacceptable change in the perturbative field equations.

Fortunately,  we can instead use integration by parts, which is guaranteed to  exactly preserve   the numerical value 
of $\widetilde{\mathcal{L}}_\text{NLO}$  \textit{and} the perturbative field equations. 
With the help of identity~(\ref{eq:identity}), we perform the replacement 
\begin{equation}
\widetilde{g}_5(X)  X_t  \nabla^2 \theta \rightarrow -\widetilde{g}_5(X) (\bm{\nabla} X )^2 
+ \widetilde{G}_5(X) \nabla^2 V_\text{ext} + 
\frac{1}{m} \widetilde{G}_5(X) \left[(\partial_i \partial_j \theta)^2 - (\nabla^2 \theta )^2 \right]  , 
\end{equation}
where $\widetilde{G}_5'(X) = \widetilde{g}_5(X)$.
Thus, the reduced effective Lagrangian becomes 
\begin{equation}\label{eq:red}
\mathcal{L}=P(X)+  f_1(X) (\bm{\nabla} X )^2 + 
 f_2(X)(\nabla^2 \theta)^2 +  f_3(X)  (\partial_i \partial_j \theta)^2 + 
 f_4 (X)  \nabla^2 V_\text{ext}, 
\end{equation}
where the coefficient functions are given by 
\begin{subequations}
\label{eq:fi}
\begin{eqnarray}
 f_1(X) &=& g_1(X) -\widetilde{G}_5'(X) , \\
 f_2(X) &=& g_2(X) - g_4(X) \left(\frac{P'(X)}{m P''(X)}\right)^2-
  \frac{1}{m}\widetilde{G}_5(X), \\  
 f_3(X) &=&g_3(X) + \frac{1}{m}\widetilde{G}_5(X) , \\ 
 f_4(X)&=& \widetilde{G}_5(X) ,
\end{eqnarray}
\end{subequations}
with 
\begin{equation}
\widetilde{G}_5'(X) = g_5(X) - 2 g_4(X) \frac{P'(X)}{m P''(X)} .
\end{equation} 
The constant of integration in $\widetilde{G}_5(X)$ is irrelevant, since it enters  the effective Lagrangian 
as the coefficient of a total divergence, as follows from  
Eq.~(\ref{eq:identity}) when $G(X)$ is a constant. 

Note that the term proportional $(\partial_i \partial_j \theta)^2 $, which has not been considered previously in the literature, is an essential ingredient of the effective Lagrangian and  can not be eliminated without introducing unacceptable changes in the NLO field equations: Although we could use Eq.~(\ref{eq:identity}) to eliminate this term in favour of the other three invariants, in doing so we would reintroduce  the term proportional to $X_t  \nabla^2\theta$ which, as we have seen, should not be eliminated through the use of the LO field equations.
 
An alternative way to understand this issue is by noting that using  the LO field equation in Eq.~(\ref{eq:arg}) would be  equivalent, up to $O(p^4)$ terms arising from the second variation of $P(X)$, to performing the field redefinition\footnote{We are indebted to the authors of Ref.~\cite{Son} for this observation.} $\tilde\theta=\theta+\delta\theta$
\beq\label{eq:redef}
\delta\theta=\frac{ \tilde{g}_5(X)}{P''(X)}\nabla^2 \theta
\eeq
The variation of $P(X)$ under this redefinition cancels the last term in Eq.~(\ref{eq:int}) and we are left with an effective NLO Lagrangian depending  on only three coefficient functions. Since a field redefinition can  change neither the value of the action evaluated at its extrema (the classical action) nor the result of any functional integration (the generating functional), it is clear that, for some applications, one can use an efffective NLO Lagrangian with only three coefficient functions. 
However, the field redefinition~(\ref{eq:redef}) involves a change  of $O(p^2)$, making $\theta$ and $\tilde\theta$ non-equivalent at NLO. Thus, with hydrodynamical applications in view where the field $\theta$ has a concrete physical meaning ---through the relation to the superfluid velocity $\bm{v}_s =  m^{-1} \bm{\nabla} \theta$--- one is forced to keep all four NLO  coefficient functions in~(\ref{eq:red}). Only this way can we preserve, at NLO,  the interpretation of $\theta$ as ($1/2$) the phase of the condensate, the canonical structure of the theory and the perturbative field equations. Note that the situation is different with   the use of the LO field equation to eliminate the term proportinal to $X_t^2$ in Eq.~(\ref{eq:quad}), which is equivalent to a field redefinition with
\beq
\delta\theta=\frac{ g_4(X)}{P''(X)}\left(X_t  +\frac{P'(X)}{m P''(X)} \nabla^2 \theta\right)
\eeq
In this case the difference between $\theta$ and $\tilde\theta$ is of $O(p^4)$ and they are equivalent at NLO.


\section{Effective parameters in the mean-field approximation} 
\label{sec:para} 

The effective theory of the previous Section contains a set of functions  depending on the chemical potential and the scattering length  once  
$\Delta_0(\mu) $ is inserted. 
Calculating these  functions in the mean field approximation is relatively straightforward but tedious and, as explained in Appendix~A,   all the integrals  (\ref{eq:int}) are computable in closed form in terms of complete elliptic integrals of the first and second kinds;  the results are given in this Section and in Appendix~\ref{sec:deriv}. 
Other physical quantities,  such as the length for pair correlation, have been expressed in terms of elliptic integrals in Ref.~\cite{Marini}, but here we will focus 
on the thermodynamical potential and the coefficients $f_i(X)$. 
Henceforth the particle density is written as $\langle n\rangle \equiv k_F^3/3\pi^2$, where $k_F$ is the Fermi wave vector.   

\subsection{Leading order results}

Here, we briefly collect the  results for the pressure and the static, zero-momentum susceptibilities in the mean field approximation. 
We start with the one-loop thermodynamic potential~\cite{Diener}  
\begin{equation}
\frac{\Omega}{V} = -\frac{m \Delta_0^2}{4 \pi a}  - 
   \int \frac{d^3 k}{(2 \pi)^3} \left( E_{\bm{k}}  - \epsilon_{\bm{k}} + \mu  - \frac{1}{2} \frac{\Delta_0^2}{\epsilon_{\bm{k}}} \right) , 
\end{equation}
which, as mentioned above, can be differentiated to yield 
Eqs.~(\ref{eq:nn}), (\ref{eq:n1}) and (\ref{eq:11}) in the same approximation. 
The integral can be done anlytically, giving
\begin{eqnarray}
\label{eq:thermo}
\frac{\Omega}{V}  &=& -\frac{m \Delta_0^2}{4 \pi a} - \frac{4 m^{3/2} |\mu|^{5/2}}{15 \pi^2} \left[ 
  (1-3 \alpha^2) \sqrt{-\text{sign}(\mu) + \sqrt{1+\alpha^2}} \, E(-\gamma) \right. \nonumber \\ 
 && \left. +\frac{\text{sign}(\mu) (1+\alpha^2)  + ( 3 \alpha^2-1) \sqrt{1 + \alpha^2} }{\sqrt{-\text{sign}(\mu) + \sqrt{1+\alpha^2}}}  
K(-\gamma) \right] ,
\end{eqnarray}
where $K(n)$,  $E(n)$  are  the complete elliptic integrals of the first and second kind respectively (see Appendix~A for details and notation), and  $\alpha^2 = \Delta_0^2/\mu^2$.  
The parameter $\gamma$ is 
\begin{equation}
\gamma = \frac{\left[ \sqrt{1+\alpha^2} + \text{sign}(\mu)\right]^2}{\alpha^2} .
\end{equation}

The susceptibilities are given by
\begin{eqnarray}
 \chi_{nn}(0)&=&   
\frac{m^{3/2} |\mu|^{1/2}}{\pi^2}  \sqrt{-\text{sign}(\mu) + \sqrt{1+\alpha^2}} 
\left[ -E(-\gamma)+K(-\gamma) \right] , \\ 
 \chi_{n1}(0)&=&    
-\frac{m^{3/2} \Delta_0} {\pi^2 |\mu|^{1/2}}\frac{1}{  \sqrt{-\text{sign}(\mu) + \sqrt{1+\alpha^2}} } K(-\gamma)  , \\ 
\chi_{1 1}(0)&=&   
\frac{m^{3/2} |\mu|^{1/2}}{\pi^2}\frac{1} {\alpha^2 \sqrt{1+\alpha^2}}  \sqrt{-\text{sign}(\mu) + \sqrt{1+\alpha^2}}
\left[ 3 \alpha^2 \sqrt{1+\alpha^2} E(-\gamma) \right.  \nonumber  \\ 
&& \left. -  \left( 2\,\text{sign}(\mu) (1+\alpha^2) + (2+3 \alpha^2) \sqrt{1+\alpha^2} \right) K(-\gamma) \right] . 
 \end{eqnarray}
The expression for the thermodynamic potential~(\ref{eq:thermo}),  the gap Eq.~(\ref{eq:gap}) and the number equation 
\begin{equation}
\langle n \rangle = -\frac{1}{V} \frac{\partial \Omega}{\partial \mu} 
\end{equation}
summarize the properties of the crossover at $T=0$ in the one-loop approximation. 
From these one can find the quantities $\Delta_0/\epsilon_F$ and $\mu/\epsilon_F$ as functions of the parameter $(k_F a)^{-1}$, where $\epsilon_F = k_F^2/2 m$ is the Fermi energy. The energy density of the ground state  is given by $\epsilon = \mu \langle n \rangle + V^{-1}\Omega$. 

Next we collect  results, some of then well known, which follow  easily from Eq.~(\ref{eq:thermo}) for specific ranges of the parameter $(k_F a)^{-1}$. 

\subsubsection{Near unitarity}
Here we present  the results for the gap parameter, the chemical potential and the ground state energy per particle near unitarity, expressed as power series in 
$(k_F a)^{-1}$:
\begin{eqnarray}
\frac{\Delta_0}{\epsilon_F} &=& 0.6864 + \frac{0.6368}{k_F a} + 
 \frac{0.0959}{(k_F a)^2} + \dots , \\
\frac{\mu}{\epsilon_F} &=& 0.5906 - \frac{0.7401}{k_F a} - 
 \frac{0.5150}{(k_F a)^2} + \dots , \\  
\frac{\varepsilon}{\langle n \rangle} &=& \frac{3 \epsilon_F}{5} \left(0.5906 - \frac{0.9251}{k_F a} - 
 \frac{0.8582}{(k_F a)^2} + \dots \right) .
\end{eqnarray} 
The numerical coefficients have been  obtained by simultaneous  power series solution of the gap and number equations using the  
analytic solution~(\ref{eq:thermo})  for $\Omega$. It is also  easy to write a few terms for the pressure at large scattering length
\begin{equation}
\label{eq:pumu}
P(\mu) = 0.0842\, m^{3/2} \mu^{5/2} + 0.1075\frac{m \mu^2}{a} + 
0.1274\frac{m^{1/2} \mu^{3/2}}{a^2} + 0.1006\frac{\mu}{a^3} + \ldots
\end{equation}

\subsubsection{BCS limit}
In the BCS limit $(k_F a)^{-1} \rightarrow-\infty$. By using the following asymptotic expansions for the complete elliptic integrals, valid when $z \rightarrow -\infty$
\begin{eqnarray}
K(z) &\sim& \frac{\ln(-16 z)}{2(-z)^{1/2}} + 
            \frac{2-\ln(-16 z)}{8(-z)^{3/2}} + 
           \mathcal{O}\left((-z)^{-5/2}\ln(-z)\right), \\ 
E(z) &\sim & (-z)^{1/2} + \frac{1+\ln(-16 z)}{4(-z)^{1/2}} +
           \frac{3-2\ln(-16 z)}{64(-z)^{3/2}}+ 
           \mathcal{O}\left((-z)^{-5/2}\ln(-z)\right),
\end{eqnarray}
we obtain the solution to the gap equation $\alpha = 8 e^{-2} \exp({\pi/2 \sqrt{2 m \mu} a)})$. This produces the pressure
\begin{equation}
\label{eq:statebcs}
P(\mu)  =  \frac{2^{5/2} m^{3/2} \mu^{5/2}}{15  \pi^2 } 
  \left(1 + 60 \exp(-4+\pi a^{-1} (2 m \mu)^{-1/2}) + \ldots  \right) , 
\end{equation}
and 
\begin{eqnarray}
\frac{\Delta_0}{\epsilon_F} &=& 8 e^{-2} e^{\pi/2 k_F a}, \\
\frac{\mu}{\epsilon_F} &=& 1 + \frac{8 \pi e^{-4} e^{\pi/k_F a}}{k_F a}+ \dots = 
  1 + \frac{\Delta_0^2}{4 \epsilon_F^2} 
  \left[2 + \ln\left(\frac{\Delta_0}{8 \epsilon_F} \right) \right]+ \dots 
\end{eqnarray}

We find that the ground state energy per particle and the fermion density are  given by 
\begin{eqnarray}
\frac{\epsilon}{\langle n \rangle} &=& \frac{3\epsilon_F}{5} - \frac{3 \Delta_0^2}{8 \epsilon_F} + \dots, \\ 
\langle n \rangle  &=& \frac{k_F^3}{3 \pi^2}\left[
  1 + \frac{3 \Delta_0^2}{16 \epsilon_F^2}\left(1- 
  2 \ln\left(\frac{\Delta_0}{8 \epsilon_F} \right)\right) + \ldots \right].
\end{eqnarray}

\subsubsection{BEC limit}
The other extreme regime, the BEC limit $(k_\text{F} a)^{-1} \rightarrow \infty$, is  obtained when $\mu$ is close to $-1/2 m a^2$.  
In this regime $|\alpha|$ is small and we can use the Maclaurin series for the elliptic integrals. The thermodynamic potential is thus
\begin{equation}
\frac{\Omega}{V} = -\frac{m \Delta_0^2}{4 \pi a} + 
\frac{m^{3/2} (-\mu)^{1/2} \Delta_0^2}{2 \pi \sqrt{2}} + 
\frac{m^{3/2} \Delta_0^4}{64\pi \sqrt{2} (-\mu)^{3/2}} + \dots
\end{equation} 
and one finds
\begin{eqnarray}
\mu &=& -\frac{1}{2 m a^2} + \delta \mu = 
\epsilon_F \left(-\frac{1}{(k_F a)^2} + \frac{2 k_F a}{3 \pi} + \dots \right) , \\ 
\Delta_0 &=& 
\epsilon_F \frac{4}{(3\pi k_F a)^{1/2}} + \dots,  \\ 
\frac{\epsilon}{\langle n \rangle} &=&  
\epsilon_F \left(-\frac{1}{(k_F a)^2} + \frac{k_F a}{3 \pi} + \dots \right) .
\end{eqnarray}
The solution of the gap equation yields a power series in $m a^2 \delta\mu$ for $\Delta_0 $ 
\begin{equation}
\label{eq:BECdelta}
\Delta_0 = \frac{2 \sqrt{\delta\mu}}{a \sqrt{m}}\left(1 + \frac{5}{8} m a^2 \delta\mu + 
O((m a^2 \delta\mu)^2) \right), 
\end{equation}
which can be used to write  the  pressure as  
\begin{equation}
P = \frac{m \delta \mu^2}{2\pi a}  + \frac{m^2 a\, \delta\mu^3}{4 \pi} + \ldots
\end{equation}

Eliminating the chemical potential leads to a pressure 
$P(n)= (2 m)^{-1} \pi a n^2 +\ldots$. Noting that $m_B = 2 m$ and $n_B = n/2$, 
a comparison with the pressure of a weakly interacting Bose gas in the lowest-order approximation 
$P(n_B)= 2 \pi a_B m_B^{-1}  n_B^2$, yields the mean-field result for the scattering length between bosons,  $a_B = 2 a$.  
But, as Diener, Sensarma and Randeria have recently shown~\cite{Diener}, these mean-field results in the BEC limit are poor approximations to the results which are 
obtained when the contribution from the quantum fluctuations is included in the computation.


\subsection{Next-to-leading order results in the one-loop approximation}

The derivatives of the one-loop response functions at $Q=0$ are collected in Appendix~\ref{sec:deriv}. 
From these expressions we can  determine the  coefficient functions $f_i(X)$ in the effective NLO Lagrangian
\begin{equation}
\label{eq:f14}
\mathcal{L}_\text{NLO}= f_1(X) (\bm{\nabla} X )^2 + 
 f_2(X)(\nabla^2 \theta)^2 +  f_3(X)  (\partial_i \partial_j \theta)^2 + 
 f_4 (X)  \nabla^2 V_\text{ext}. 
\end{equation}

\subsubsection{Unitarity}
 It is possible to express the complete elliptic integrals in the unitary limit through two useful formulas  
 obtained by  simultaneous solution of  the  pressure $P= c_0 m^{3/2} \mu^{5/2}$ and  gap equations. These are given by  
\begin{eqnarray}
K(-\gamma)&=&\frac{15 \pi^2 \sqrt{-1 + \sqrt{1+\alpha^2} }}{4 (1+\alpha^2)}\, c_0 , \\ 
E(-\gamma)&=&\frac{15 \pi^2}{4 \sqrt{1+\alpha^2}\sqrt{-1 + \sqrt{1+\alpha^2} }}\, c_0. 
\end{eqnarray}
Substituting these expressions into the derivatives of Appendix C produces the  
remarkable result that   $\widetilde{G}_5'(\mu) \eq 0$.
This is due to the fact that, when the coupling $g \rightarrow \infty$,  the expression for this quantity is proportional to  the gap equation, 
 which gives rise to the exact cancellation of $\widetilde{G}_5'(\mu)$. 
 The complete list of  coefficient functions that we find is
\begin{subequations} 
\begin{eqnarray}
f_1(\mu) &=& -\frac{35 c_0}{192}  \frac{m^{1/2}}{\mu^{1/2}} \simeq 
-0.0153 \frac{m^{1/2}}{\mu^{1/2}}, \\ 
f_2(\mu) &=& -\frac{c_0}{18 \alpha(\mu)^2} \frac{\mu^{1/2}}{m^{1/2}} \simeq  
 -0.0035  \frac{\mu^{1/2}}{m^{1/2}} , \\
f_3(\mu) &=& \frac{c_0}{6 \alpha(\mu)^2} \frac{\mu^{1/2}}{m^{1/2}} \simeq  
 0.0104  \frac{\mu^{1/2}}{m^{1/2}} , \\ 
f_4(\mu) &=& 0,  
\end{eqnarray}
\end{subequations} 
where the numerical values are obtained by substitution of the one-loop numerical values 
$\alpha(\mu) \simeq 1.1622$ and $c_0  \simeq 0.0842$. 
 Noting that  $f_3/f_2 = -3$, our result for the NLO effective Lagrangian at unitarity can be written as
 \beq
 \label{eq:c1c2}
 \mathcal{L}_\text{NLO} =  c_1 m^{1/2}  X^{-1/2} (\bm{\nabla} X)^2  
+ c_2   m^{-1/2} X^{1/2}\left[ (\nabla^2\theta)^2 -3 (\partial_i \partial_j \theta)^2 \right], 
 \eeq
 where $c_1\simeq -0.0153$ and $c_2\simeq -0.0035$. This is one of the main results in this paper. 
 
 One might wonder whether the cancellation of the $f_4$ coefficient and the ratio $f_3/f_2 = -3$ found here are mere accidents of the mean field approximation.  Actually, these are exact consequences of the conformal invariance displayed by the system at the unitarity limit. In other words, any approximation scheme that respects conformal invariance must necessarily yield a result of the form given by Eq.~(\ref{eq:c1c2}). This is explained in Appendix~\ref{sec:symm}, where the reason for the discrepancy with the form of the effective action given in~\cite{Son} is also analyzed. 

 \subsubsection{Near unitarity}
 The following expansions are valid for the NLO coefficient functions near unitarity 
 \begin{subequations} 
 \begin{eqnarray}
 f_1(\mu) &=&  -\frac{0.0153 m^{1/2}}{\mu^{1/2}} +  \frac{ 0.0014 }{a \mu} 
 - \frac{0.0016}{a^2 m^{1/2} \mu^{3 /2}} + \ldots,  \\
  f_2(\mu) &=&  -\frac{0.0035 \mu^{1/2}}{m^{1/2}} +
   \left[0.0023 +  0.0022 \ln \left(\frac{\mu}{\mu_0} \right) \right]\frac{1}{a m} \nonumber \\
 &&+ \frac{0.0004 }{a^2 m^{3/2} \mu^{1 /2}}  + \frac{0.0022 }{a^3 m^{2} \mu}  + \ldots,  \\ 
 f_3(\mu) &=&  \frac{0.0104 \mu^{1/2}}{m^{1/2}} -
   \left[0.0113 +  0.0022 \ln \left(\frac{\mu}{\mu_0} \right) \right]\frac{1}{a m} \nonumber \\
 &&+ \frac{0.0097 }{a^2 m^{3/2} \mu^{1 /2}}  - \frac{0.0094 }{a^3 m^{2} \mu}  + \ldots, \\
  f_4(\mu) &=&  - \frac{0.0022}{a} \ln \left(\frac{\mu}{\mu_0} \right)  
 - \frac{0.0037 }{a^2 m^{1/2} \mu^{1 /2}}  + \frac{0.0004 }{a^3 m \mu}  + \ldots,
 \end{eqnarray}
\end{subequations} 
where $\mu_0$ is an arbitrary scale ($\ln\mu_0$  is multiplied by a total divergence).

\subsubsection{BCS limit}
 
 The determination of the leading behavior of the coefficient functions in the BCS limit is more involved. 
By using the asymptotic expansions for the 
complete elliptic integrals and the perturbative solution of the gap equation   $\alpha(\mu)\approx 8 e^{-2} \exp(\pi/2 a \sqrt{2 m \mu} )$, one finds the lowest-order approximation 
\begin{equation}
\frac{g \chi_{n1}(0)}{2+g \chi_{11}(0)} = 
\frac{\alpha(\mu)}{2} \ln \left(\frac{\alpha(\mu)}{8} \right) + 
O(\alpha^3 (\ln \alpha)^2), 
\end{equation}
to be used in the coefficients of some derivatives in the asymptotic expressions for the $b_i$ coefficients.    
The pressure ~(\ref{eq:statebcs}) gives
\begin{equation}
  \frac{P'(\mu)}{m P''(\mu)} =  \frac{2 \mu}{3 m} - \frac{\mu}{6 m} \alpha(\mu)^2
  \ln\left(\frac{\alpha(\mu)}{8} \right) \left[ 2 + \ln\left(\frac{\alpha(\mu)}{8} \right)\right]+ O(\alpha^4 (\ln \alpha)^4),
\end{equation}
and the substitution of these results into Eqs.~(\ref{eq:fi})  produces
\begin{equation}
\widetilde{G}_5'(\mu) = \frac{\left(\ln[\alpha(\mu)/8] \right)^2}{36 \sqrt{2} \pi^2}  \frac{m^{1/2}}{\mu^{1/2}} + \ldots 
=\frac{1}{288 \sqrt{2}\,  m^{1/2} a^2 \mu^{3/2}} +\ldots,  
\end{equation}
and 
\begin{subequations} 
\label{eq:nlobcs}
\begin{eqnarray}
 f_1(\mu) &=&-\frac{\left(\ln[\alpha(\mu)/8] \right)^2}{24 \sqrt{2} \pi^2}  \frac{m^{1/2}}{\mu^{1/2}} + \ldots 
   =-\frac{1}{192 \sqrt{2}\,  m^{1/2} a^2 \mu^{3/2}} +\ldots , \\    
 f_2(\mu) &=&-\frac{2^{3/2}}{135 \pi^2 \alpha(\mu)^2} \frac{\mu^{1/2}}{m^{1/2}}+\ldots = 
  -\frac{\exp(4- \pi a^{-1} (2 m \mu)^{-1/2})}{2160 \sqrt{2}\, \pi^2 }
  \frac{\mu^{1/2}}{m^{1/2}} + \ldots    ,  \\ 
 f_3(\mu) &=&\frac{2^{3/2}}{45 \pi^2 \alpha(\mu)^2} \frac{\mu^{1/2}}{m^{1/2}}+\ldots = 
  \frac{\exp(4- \pi a^{-1} (2 m \mu)^{-1/2})}{720 \sqrt{2}\, \pi^2 }
  \frac{\mu^{1/2}}{m^{1/2}} + \ldots , \\  
 f_4(\mu) &=& -\frac{1}{144 \sqrt{2}\,  m^{1/2} a^2 \mu^{1/2}} +\ldots
\end{eqnarray} 
\end{subequations}
It is worth noting that $f_4/(m f_{2,3}) = O(\alpha^2 (\ln \alpha)^2)$, 
which shows that $f_4$ can be safely neglected. 
Hence, an important feature of this regime is the dominance of the $f_2$ and $f_3$ terms in the 
next-to-leading order effective Lagrangian. 
The coefficients $b_4$ and $b_5$ are at the origin of the leading behavior of $f_{2,3}$,
while $b_3$ and $b_4$ govern the expression for $f_1$.  

Note  that these expressions  can not be trusted   for  $k_F a$ arbitrarily small, because then 
$\Delta_0 \rightarrow  0$, and 
the  condition for  the validity of the derivative expansion $|\omega| < 2 \Delta_0$  cannot be satisfied. 
Indeed, when $a \rightarrow 0^-$ the Goldstone mode can hardly be considered a propagating mode due  the arbitrarily small two-particle states threshold, and this renders this effective field description meaningless.

\subsubsection{BEC limit} 
The BEC limit can be obtained by substituting  the gap parameter~(\ref{eq:BECdelta}) into the appropriate derivatives of the response functions and  then expanding in powers of $\delta\mu$.   We obtain the following leading behavior for the coefficient functions
\begin{subequations} 
\begin{eqnarray}
 \widetilde{G}_5'(\mu) &=& -\frac{m a}{48 \pi}  + O(\delta \mu), \\
 \label{eq:f1bec}
 f_1(\mu) &=&-\frac{1}{8\pi m a^3 \Delta_0(\mu)^2}+O(\delta \mu^0) = 
   -\frac{m a}{16 \pi(1 + 2 m a^2 \mu)} +O(\delta \mu^0), \\ 
 f_2(\mu) &=&  -\frac{1}{96 \pi m  a } + O(\delta \mu^2),  \\ 
 f_3(\mu) &=&  \frac{1}{96 \pi m a } + O(\delta \mu^2),  \\  
 f_4(\mu) &=& -\frac{m a \mu}{48 \pi} + O(\delta \mu^2) .
 \end{eqnarray}
\end{subequations} 
These results reveal that, in this limit,  $\widetilde{G}_5'(\mu)/f_1(\mu) = O(m a^2 \delta\mu)$,  
and the leading derivative contribution comes from the $f_1$  term of the effective Lagrangian. 
Now, the coefficient $b_3$ is at the origin of the leading behavior of $f_1$.
Note that the constant coefficients  $f_2$ and $f_3$ play no role in this limit, because an integration by parts  produces their mutual  cancellation.


\section{Density functional in the unitary limit. Relation to other approaches}
\label{sec:functional}

From the effective Lagrangian one can easily derive the energy density  $\mathcal{E}$ depending on the number density and spatial derivatives of the Goldstone mode. The corresponding first-order equations are the continuity equation and the London equation for $\theta$. 
Here we outline the computation of $\mathcal{E}$. 

Since s 
$\mathcal{L}^\text{eff} = n (-\partial_t \theta -( \nabla \theta)^2/2m) + \ldots$ one sees that the number density $n$ is conjugate to $-\theta$, and the  
energy density is given by 
\begin{equation}
\mathcal{E} = -n \partial_t \theta + \mu n - \mathcal{L} ,  
\end{equation} 
where $\partial_t \theta$ is determined in terms of  $n$,  $\nabla \theta$ and $V_\text{ext}$ by assuming a derivative expansion for 
$\partial\mathcal{L}/\partial(\partial_t \theta)=-n$.
From the effective Lagrangian at unitarity obtained in the previous Section
\beq
\label{eq:lunit}
\mathcal{L} = c_0 m^{3/2} X^{5/2} +  c_1 m^{1/2}  X^{-1/2} (\bm{\nabla} X)^2  
+ c_2   m^{-1/2} X^{1/2}\left[ (\nabla^2\theta)^2 -3 (\partial_i \partial_j \theta)^2 \right], 
 \eeq
one finds the energy density 
\beqa
\label{eq:energy}
\mathcal{E} &=& n V_\text{ext}  + 
\frac{3 \cdot 2^{2/3}}{5^{5/3} c_0^{2/3} m} n^{5/3} + 
 n \frac{(\bm{\nabla}\theta)^2}{2 m} -  
 \frac{8 c_1}{45 c_0 m} \frac{(\bm{\nabla}n)^2}{n}\non\\
 &&  -\frac{2^{1/3} c_2 }{5^{1/3}c_0^{1/3} m} n^{1/3}  
  \left[ (\nabla^2\theta)^2  - 3 (\partial_i \partial_j \theta)^2\right] .
\eeqa 
Note that the next-to-leading order contribution to $\mathcal{E} $ is exactly 
$-\mathcal{L}_\text{NLO}$ if $\partial_t \theta$  is replaced by 
\begin{equation}
\label{eq:london}
\partial_t \theta =\mu -V_\text{ext} - \frac{2^{2/3} n^{2/3}}{5^{2/3} c_0^{2/3} m}  - \frac{(\bm{\nabla}\theta)^2}{2 m} , 
\end{equation}
which is the solution of $\partial\mathcal{L}_\text{LO}/\partial(\partial_t \theta)=-n$.

The variation of the energy functional $H[n,\theta] = \int d^3 x \mathcal{E}$ with respect to  $\theta$ yields  the continuity equation for the particle number 
\begin{equation}
\partial_t n = \frac{\delta H}{\delta \theta(x)} = - \bm{\nabla} \cdot \frac{\partial \mathcal{E}}{\partial( \bm{\nabla}  \theta(x))} + \ldots = 
-\bm{\nabla} \cdot\left( n \frac{\bm{\nabla}\theta}{m} + \ldots \right) , 
\end{equation}
and the hydrodynamic equation for $\theta$ is given by 
\begin{equation}
\partial_t \theta =  -\frac{\delta H}{\delta n(x)}    .
\end{equation}

These equations describe the irrotational hydrodynamics of the superfluid at zero temperature.
The equilibrium state in the presence of an external potential corresponds to 
$\theta = -\mu_0 t + \text{cons}$, which gives a stationary density profile and zero superfluid velocity,  
$\bm{\nabla}\theta =0$. 
The equilibrium particle density  satisfies the condition   
\begin{equation}
\label{eq:equil}
 \mu_0 = V_\text{ext} +  \frac{2^{2/3}}{5^{2/3} c_0^{2/3} m} n^{2/3}  + 
 \frac{32 c_1}{45 c_0 m} \frac{\nabla^2 (\sqrt{n})}{\sqrt{n}}  , 
\end{equation}
where $\mu_0$ is the chemical potential. 
Thus the leading behavior of $n$, which is obtained by dropping all the $c_i$, is given by a Thomas-Fermi approximation, 
whereas  $c_1$ determines the quantum kinetic energy correction and, as shown in Appendix~\ref{sec:symm}, the  last  term  is proportional to the square of  the traceless  strain rate tensor.

In writing Eq.~(\ref{eq:lunit}) we have used the results of the previous Section, namely $f_4=0$ and $f_3/f_2 = -3$. It is worth mentioning that, had we found a nonvanishing value for $f_4$, the effective Lagrangian~(\ref{eq:lunit}) would have contained an additional term proportional to 
$m^{1/2} X^{1/2}\nabla^2 V_\text{ext}$, giving rise to a term  proportional to $n^{-2/3} \nabla^2 V_\text{ext}$ in~(\ref{eq:equil}). Such a term would be incompatible with the  boundary condition for $n$ at infinity, due to the negative power of $n$. If we consider 
 for instance the isotropic harmonic trap, where $\nabla^2 V_\text{ext}$ is  a constant, it is not possible to satisfy the 
asymptotics $n \rightarrow 0$ as $r  \rightarrow \infty$.   
But, as mentioned above and shown explicitly in Appendix~B, the condition $f_4=0$ is an exact  consequence of the conformal invariance of the field equations and the problem with the boundary conditions does not arise.

Very recently, Rupak and Sch\"afer~\cite{Rupak} have derived an energy density functional using  an epsilon expansion around $d=4-\varepsilon$ spatial dimensions. 
They find an expression depending only on $n$ given by 
\begin{equation}
\mathcal{E}_\mathrm{Rupak, Sch\ddot{a}ffer}= V_\text{ext} n  +1.364 \frac{n^{5/3}}{m}+ 
                             0.032 \frac{(\bm{\nabla}n)^2}{m n} + O(\nabla^4 n)  , 
\end{equation}
which follows from their value of $c_1 \simeq -0.0209$ up to $O(\varepsilon \ln\varepsilon)$, 
together with $\xi \simeq 0.4754$ 
(or $c_0 \simeq 0.1166$) in the same approximation~\cite{Nishida}.
Our result for $\mathcal{E}$ is obtained by inserting the mean field results  $c_0 = 2^{5/2}/(15 \pi^2 \xi^{3/2}) = 0.0842$, $c_1/c_0 = -35/192$ and   
$c_2 = -0.0035$  into Eq.~(\ref{eq:energy}). 
We find that  the energy density takes the form
\begin{eqnarray}
\mathcal{E} &=& V_\text{ext}  n  +1.6956 \frac{n^{5/3}}{m}
                              + n \frac{(\bm{\nabla}\theta)^2}{2 m} + 
                              0.0324 \frac{(\bm{\nabla}n)^2}{m n}  \nonumber \\
                              && + 
                             0.0059 \frac{n^{1/3}}{m} \left( (\nabla^2 \theta)^2 -3 (\partial_i \partial_j \theta)^2 \right). 
\end{eqnarray}
It is remarkable the agreement between the terms corresponding to the quantum kinetic energy or quantum pressure, although they have been computed using two very different approaches.  
Although the values  for the individual coefficients $c_0$ and $c_1$ show differences of the order of $30\%$,  surprisingly these differences cancel  in the ratio $c_1/c_0$  giving the quantum pressure.

It is interesting to consider the expression for the energy density in the BEC limit which follows from the effective Lagrangian. 
It can be written in the form
\begin{equation}
\mathcal{E}_\text{BEC}  = n V_\text{ext}  + 
\frac{\pi a}{2 m} n^2 + 
 n \frac{(\bm{\nabla}\theta)^2}{2 m} +
 \frac{1}{32} \frac{(\bm{\nabla}n)^2}{m n} +\frac{1}{96 \pi a } (-1 + 2 \pi a^3  n)   \nabla^2 V_\text{ext} , 
\end{equation}
where the most important derivative term is determined by the expression for $f_1(X)$ given in Eq.~(\ref{eq:f1bec}). 
If we make the replacements $n \rightarrow 2 n_B$,  $m \rightarrow m_B/2$ and $a \rightarrow a_B/2$, this expression 
becomes
\begin{equation}
\mathcal{E}_\text{BEC}  = n_B  (2 V_\text{ext})
+\frac{4\pi a_B}{2 m_B} n_B^2 + 
 n_B \frac{(\bm{\nabla}(2 \theta))^2}{2 m_B} +
 \frac{1}{2 m_B} (\bm{\nabla}\sqrt{n_B})^2+\ldots ,
\end{equation} 
and one recovers the correct quantum pressure in the last term. 
Thus, the derivative part of the effective Lagrangian in the BEC limit  fits in with the hydrodynamic description 
of a  superfluid at zero temperature
with bosonic constituents of mass $2 m$.  
It is worth pointing out the excellent numerical agreement between the coefficient $1/32\simeq 0.031$ in the BEC limit  and  
the corresponding coefficient $7/216\simeq 0.032$ in the unitary limit. This is probably a reflection of the fact that, in the $d=4$ limit  about which  the epsilon expansion is taken, the fermion-fermion scattering amplitude is saturated by the propagator of a boson with mass $2m$. 

Finally, in order to check the quality of the above NLO results in the BCS limit,
a  comparison  with the predictions from the approach of 
Furnstahl \textit{et al.}~\cite{Furnstahl} would be very insightful.


\section{Conclusion}
\label{sec:end}
In this paper we have considered the derivative expansion of the effective action for the Goldstone field of a nonrelativistic superfluid Fermi gas at zero temperature in the whole BCS-BEC crossover.  Based on  the pioneering analysis of  symmetries in Ref.~\cite{Son}, we have shown that the NLO action  can be given in terms of the four coefficient functions in 
\begin{equation}
\label{eq:novel}
\mathcal{L}=P(X)+  f_1(X) (\bm{\nabla} X )^2 + 
 f_2(X)(\nabla^2 \theta)^2 +  f_3(X)  (\partial_i \partial_j \theta)^2 + 
 f_4 (X)  \nabla^2 V_\text{ext}\;, 
\end{equation}
and we have given the precise relationships ---see Eqs.~(\ref{eq:fi})--- between  these functions and the derivatives of the response functions.
It  turns out that the computation of the NLO action  relies also on the ratio of a pair of three-point functions, whose value we have determined by a calculation  at the one-loop level. 

An important step towards Eqs.~(\ref{eq:fi}) has been the reduction of the initial Lagrangian~(\ref{eq:lag6}) to its  final form~(\ref{eq:novel}) without higher-order time derivatives. 
In this regard, we have shown how the proper use of the LO field equations to eliminate higher-order time derivatives insures the consistency of the reduction process. As novel consequences of this critical analysis, we find  the presence of a term proportional to  $(\partial_i \partial_j \theta)^2$ in~(\ref{eq:novel})  and the form of this action in the unitary limit
\beq
\label{eq:novel2}
\mathcal{L} = c_0 m^{3/2} X^{5/2} +  c_1 m^{1/2}  X^{-1/2} (\bm{\nabla} X)^2  
+ c_2   m^{-1/2} X^{1/2}\left[ (\nabla^2\theta)^2 -3 (\partial_i \partial_j \theta)^2 \right], 
 \eeq
which, as we have shown,  is dictated by conformal invariance. In particular,  conformal invariance of the field equation 
 prevents  the existence of an explicit coupling to the external field.\footnote{There remains, of course, an implicit coupling through the dependence of $X$ on $V_\text{ext}$.}  
This Lagrangian determines uniquely the form of an energy density functional depending on the particle density $n$ and the Goldstone mode $\theta$, 
\beqa
\mathcal{E} &=& n V_\text{ext}  + 
\frac{3 \cdot 2^{2/3}}{5^{5/3} c_0^{2/3} m} n^{5/3} + 
 n \frac{(\bm{\nabla}\theta)^2}{2 m} -  
 \frac{8 c_1}{45 c_0 m} \frac{(\bm{\nabla}n)^2}{n}\non\\
 &&  -\frac{2^{1/3} c_2 }{5^{1/3}c_0^{1/3} m} n^{1/3}  
  \left[ (\nabla^2\theta)^2  - 3 (\partial_i \partial_j \theta)^2\right] .
\eeqa
It is worth mentioning that these aspects of our work do not rely on the specific approximation used to compute the 2-point functions. 
Rather, they arise as an application of effective field theory  ideas and techniques.  

By resorting to the one-loop approximation and taking into account the $\mu$ dependence of $\Delta_0$ in the mean-field approximation, 
we have also obtained analytic, closed expressions in terms of complete elliptic integrals for the coefficient functions of the NLO effective Lagrangian in the whole BCS-BEC crossover. Having closed expressions for these functions, rather than purely numerical results,  makes the analytic study of different limits feasible. We have obtained series expansions near the unitary limit, and in the extreme BCS and BEC regimes. In particular, we have determined the mean-field values for  the coefficients $c_1$ and $c_2$ at unitarity, and thus we have explicitly checked the form of the Lagrangian (\ref{eq:novel2}). In  this regard, it is interesting how the one-loop result  found for the ratio of three-point functions combines with the result for $g_4$ to yield the conformally invariant combination $X^{1/2}[(\nabla^2 \theta)^2 - 3(\partial_i\partial_j\theta)^2)]$ ---see Eqs.~(\ref{eq:g2g3}) and (\ref{eq:fi}). 
Furthermore, the good agreement in the extreme regimes with known results from other  approaches suggests that our mean field approximation can be taken as a reliable, first qualitative estimate for the coefficients of the effective theory.


\begin{acknowledgments}
It is a pleasure to thank D.T. Son and M. Wingate for useful comments on a previous version of this paper, and  I. L. Egusquiza for collaboration at the early stages of this work.
This research was supported by the Spanish Ministry of Science and Technology under Grant No. FPA2005-04823,  and the Basque Government  under Grant No. IT-357-07. 
\end{acknowledgments}



\appendix
\section{On the reduction of the integrals to canonical forms}
Here we outline the method used to avaluate all the momentum integrals in this paper in terms of elliptic integrals of the first and second kind. For the sake of concreteness, we  consider the integral for $\chi_{n 1}(0)$  
\beq
\chi_{n 1}(0) =-\int \frac{d^3 k}{(2 \pi)^3} \frac{ \Delta_0 \xi_{\bm{k}} }{E_{\bm{k}}^3}\;,
\eeq
and assume $\mu>0$. The change of variables $x=k^2/2 m\mu$ brings the integral into the following  form
\beq
\chi_{n 1}(0) =\frac{m^{3/2} \Delta_0} {\sqrt{2}\pi^2 \mu^{1/2}}\int_0^\infty d x \frac{R(x)}{y}\;,
\eeq
where $y^2=x[(x-1)^2+\a^2]$ and $R(x)$ is the rational function
\beq
R(x)=\frac{x(x-1)}{(x-1)^2+\a^2}\;.\eeq

Now, the integral of the quotient of a rational function by the square root of a cubic or a quartic polynomial is, by definition, an elliptic integral. But we still have to reduce this integral to a combination of canonical forms. The fact that the integrand has no other singularities than the branch points of $y$ implies that only elliptic integrals of the first ($F$) and second kind ($E$) can be involved. These are given by~\cite{Abram}
\beqa
F(\varphi |n)&=&\int_0^{\sin\varphi}\frac{dt}{\sqrt {(1-t^2)(1-n t^2)} }\non\\
E(\varphi |n)&=&\int_0^{\sin\varphi}dt\frac{\sqrt {1-n t^2}}{\sqrt {1-t^2 } }\;.
\eeqa
The change of variables 
\beq
t=\frac{x-\sqrt{1+\a^2}}{x+\sqrt{1+\a^2}}\;,
\eeq
brings the integration interval to $(-1,1)$ and $y$ into the canonical form
\beq
\frac{dx}{y}=\frac{\sqrt{2}}{\sqrt{-1+\sqrt{1+\a^2}}}\frac{dt}{\sqrt{(1-t^2)(1+\g t^2)}}\;,
\eeq
where
\beq
\gamma = \frac{\left[ \sqrt{1+\alpha^2} + 1\right]^2}{\alpha^2} \;.
\eeq
Finally, after  using standard   ``reduction formulae"~\cite{Abram} we obtain
\beq
\chi_{n1}(0)=    
-\frac{m^{3/2} \Delta_0} {\pi^2 \mu^{1/2}}\frac{1}{  \sqrt{-1 + \sqrt{1+\alpha^2}} } K(-\gamma) \;,
\eeq
where $ K(-\gamma)=F(\pi/2|-\gamma)$ is the complete elliptic integral of the first kind.

The evaluation of the other integrals proceeds along the same lines. In all cases we have an odd power of $E_k=\mu[(x-1)^2+\a^2]$ that combines with  $\sqrt{x}$ from the measure $d^3k$ to give an integrand of the form $R(x)/y$, where only the rational function $R(x)$ changes from case to case. As the only singularities are the branch points of $y$ ---there are no additional poles--- only complete elliptic integrals of the first and second kind can result.

\section{\label{sec:symm}Conformal invariance of the NLO Lagrangian}

In this Appendix we investigate the constraints imposed by conformal invariance on the form of the  NLO Lagrangian in the unitary limit.
 As shown in Refs.~\cite{Wise,Son}, in that limit the original fermionic action\footnote{By this we mean the fermion action before the Hubbard-Stratanovich transformation leading to Eq.~(\ref{eq:lag}) is applied. } is invariant under a special set of time-dependent transformations. Actually,  Son and Wingate~\cite{Son} have discussed the invariance properties of the action  when the system is put in a curved manifold  and an  external gauge field. In order to adapt their transformations  to our case, where the metric is euclidean and the gauge field is set to zero, we have to use a ``gauge fixed" version wich includes, besides the purely conformal transformation, ``compensating" coordinate and gauge transformations. Its infinitesimal form is given by 
\begin{eqnarray}
\delta\psi &=&  \beta''(t) \frac{i m \bm{x}^2}{4} \psi -   \frac{ \beta'(t)}{2}  \bm{x} \cdot \bm{\nabla} \psi- \beta(t) \partial_t \psi - 
 \frac{3 \beta'(t)}{4} \psi , \\ 
 \delta V_\text{ext} &=& -\beta'''(t) \frac{ m \bm{x}^2}{4}  - 
   \frac{ \beta'(t)}{2}  \bm{x} \cdot \bm{\nabla} V_\text{ext} - \beta(t) \partial_t V_\text{ext} - \beta'(t) V_\text{ext}, 
\end{eqnarray} 
where $\beta(t)$ is an arbitrary function of $t$. In the notation of Ref.~\cite{Son}, this transformation can be seen as a combination of general, gauge and  conformal transformations with parameters $\xi^k =  \beta'(t) x^k/2$, $\alpha = m \beta''(t) \bm{x}^2/4$ and $\beta(t)$, respectively. With this choice, the euclidean metric and the zero  external gauge field are untouched. 
We assume that the variation of the chemical potential is given by $\delta\mu= -\beta'(t)\mu$, which  assigns to $V_\text{ext}$ all the change in the variation $\delta(V_\text{ext} - \mu)$  under a gauge transformation.  
In the effective theory the relevant field is the Goldstone mode $\theta$, which transforms inhomogeneously according to\footnote{The Goldstone field here corresponds to 
$\mu t-\theta_\text{Son,Wingate}$ of Ref.~\cite{Son}.}
\begin{equation}
\delta \theta = \beta''(t) \frac{ m \bm{x}^2}{4} -  \frac{ \beta'(t)}{2}  \bm{x} \cdot \bm{\nabla} \theta - \beta(t) \partial_t \theta .
\end{equation}

A ``scale transformation" is a  particularly simple conformal transformation  where $\beta(t)$ is a linear function. As shown in Ref.~\cite{Son}, scale invariance alone determines the form of the functions $f_i(X)$ in the NLO Lagrangian~(\ref{eq:f14})
\beq
\mathcal{L}_\text{NLO}=c_1m^{1/2}X^{-1/2} (\bm{\nabla} X)^2+m^{-1/2}X^{1/2} \left[c_2 (\nabla^2\theta)^2 + c_3 (\partial_i \partial_j \theta)^2+
 c_4 m  \nabla^2 V_\text{ext} \right] 
\eeq
The change under a general conformal transformation is then given by 
\begin{eqnarray} 
\label{eq:deltaL}
 \delta \mathcal{L}_\text{NLO}  
 &=&  (3 c_2 +c_3)   m^{1/2}  X^{1/2} \nabla^2\theta \, \beta''(t)  - \frac{3}{2}  c_4 m^{3/2} X^{1/2} \, \beta'''(t)  \nonumber  \\ 
 &&-\partial_t\left(\mathcal{L}_\text{NLO} \beta(t)\right) - \frac{1}{2}\bm{\nabla} \cdot \left(\bm{x} \mathcal{L}_\text{NLO} \beta'(t) \right) .  
 \end{eqnarray} 
Note that, as expected,  the action is automatically invariant under  scale transformations, for which $\beta''(t)=\beta'''(t)=0$. Invariance under general conformal transformations requires
\beq
\label{eq:cons}
c_3=-3 c_2\;\;\;\; , \;\;\;\; c_4=0\;,
\eeq
while the value of $c_1$ is unrestricted. This constrains the NLO Lagrangian to the form given by Eq.~(\ref{eq:c1c2}).

Note that we could integrate the variation of the $c_4$ term in~(\ref{eq:deltaL}) 
by parts to get a contribution proportional to $\beta''(t)$. Using the LO field equations to eliminate the time derivative $\partial_t X$ would then give the following result
\beq
 \delta \mathcal{L}_\text{NLO}  = (3 c_2 + c_3 + c_4) m^{1/2}X^{1/2} \nabla^2\theta \, \beta''(t)
\eeq
up to total derivatives. This suggests that the action is conformally invariant as long as
\begin{equation}
\label{eq:weak}
c_3=-3 c_2-c_4 , 
\end{equation} 
which is a weaker constraint than~(\ref{eq:cons}). But here, as was the case with the elimination of the $\widetilde{g}_5(X) X_t  \nabla^2\theta$ term at the end of Section~\ref{sec:higher}, the use of the LO field equation is not fully legitimate. The reason is that the elimination of the time derivative $\partial_t X$ in favour of $\nabla^2\theta$ would involve dropping a term proportional to the LO field equation, namely
\beq
\frac{3 c_4}{4}  m^2 X^{-1/2} \beta''(t) \left( X_t  + \frac{2 X}{3m}  \nabla^2 \theta \right)\,.\non
\eeq
It is easy to check that the Euler-Lagrange equation arising from this term gives  a contribution to the linearized field equation proportional to 
\beq
3m\partial_t\left[\beta'''(V_\text{ext} +\partial_t\theta)\right]+10\mu \beta'''\nabla^2\theta-4\mu \beta''\nabla^2V_\text{ext},
\eeq 
which can not be neglected. 
 In other words, the field equations derived from a Lagrangian subject only to the weaker constraints~(\ref{eq:weak}) will \textit{not} be invariant under conformal transformations. Thus, the coefficient of $\beta'''(t)$ in Eq.~(\ref{eq:deltaL})  must be zero irrespective of the values of $c_2$ and $c_3$. This fact, which  seems to have been overlooked by the authors of Ref.~\cite{Son}, would explain the discrepancies with their results.\footnote{See also comments at the end of Section~\ref{sec:higher}.} 
  
  We end this Appendix by noting that the  the conformally invariant  combination \hbox{$(\nabla^2\theta)^2 -3 (\partial_i \partial_j \theta)^2$}  can be written as the square of an $l=2$  irreducible rank-two tensor of conformal dimension one
  \beq\label{eq:square}
(\nabla^2\theta)^2 -3 (\partial_i \partial_j \theta)^2=-3\left(\partial_i \partial_j \theta-\frac{1}{3}\delta_{ij}\nabla^2\theta\right)^2. 
\eeq
This allows a  natural interpretation in terms of superfluid hydrodynamics. Writing the RHS of~(\ref{eq:square})  in terms of the  superfluid velocity   $\bm{v_s} =  m^{-1} \bm{\nabla} \theta$ shows that the new invariant is proportional to the square of the traceless part of the  \textit{strain rate tensor}, also known as the  \textit{shear rate  tensor} 
\beq
\med (\partial_i v_{s,j}+\partial_j v_{s,i})-\frac{1}{3}\, \delta_{ij}\bm{\nabla}\cdot\bm{v}_s.
\eeq


\section{\label{sec:deriv}Derivatives of the one-loop response functions}

Here we give  the derivatives of the response functions used in the computation of the NLO action at the one-loop level. The results are 
\begin{eqnarray}
\frac{\partial \chi_{nn}}{\partial \omega^2}&=& -\int {d^3 k\over (2\pi)^3}  {\D_0^2\over 4 E_k^5} \non\\
&=&  
\frac{ m^{3/2}}{24 \pi^2 |\mu|^{3/2}}  \frac{1}{\alpha^2+\alpha^4}\sqrt{-\text{sign}(\mu) + \sqrt{1+\alpha^2}} 
\left[  \left(-4-3 \alpha^2\right) E(-\gamma) \right. \nonumber \\ 
 && \left. + \left(3 + 3 \alpha^2 - \text{sign}(\mu) \sqrt{1+\alpha^2}\right) K(-\gamma) \right] , \\ 
\frac{\partial \chi_{nn}}{\partial q^2}&=& \int {d^3 k\over (2\pi)^3}\left(  {3\D_0^2\xi_k\over8 m E^5_k}+ {\D_0^2(\D_0^2-4\xi_k^2) k^2\over 24m^2 E_k^7   }
\right)\non\\
&=&   
\frac{ m^{1/2}}{24 \pi^2 |\mu|^{1/2}}  \frac{1}{1+\alpha^2}\sqrt{-\text{sign}(\mu) + \sqrt{1+\alpha^2}} 
\left[  \text{sign}(\mu)  E(-\gamma) \right. \nonumber \\ 
 && \left. + \sqrt{1+\alpha^2} K(-\gamma) \right] , \\  
\frac{\partial \chi_{n1}}{\partial \omega^2}&=& -\int {d^3 k\over (2\pi)^3}{\D_0\xi_k\over 4 E_k^5} \non\\&=&   
\frac{ m^{3/2} \Delta_0}{24 \pi^2 |\mu|^{5/2}}  \frac{1}{\alpha^2+\alpha^4}\sqrt{-\text{sign}(\mu) + \sqrt{1+\alpha^2}} 
\left[  -\text{sign}(\mu) E(-\gamma) \right. \nonumber \\ 
 && \left. - \sqrt{1+\alpha^2} K(-\gamma) \right] , \\ 
 \frac{\partial \chi_{n1}}{\partial q^2}&=& \int {d^3 k\over (2\pi)^3}\left( {\D_0(2\xi_k ^2-\D_0^2)\over 8 m E_k^5} +{ \D_0\xi_k(3\D_0^2-2\xi_k^2) k^2\over 24 m^2 E_k^7}\right)\non\\&=&   
\frac{ m^{1/2} \Delta_0}{24 \pi^2 |\mu|^{3/2}}  \frac{1}{\alpha^2+\alpha^4}\sqrt{-\text{sign}(\mu) + \sqrt{1+\alpha^2}} 
\left[  \alpha^2 E(-\gamma) \right. \nonumber \\ 
 && \left. - \left(1 +  \alpha^2 + \text{sign}(\mu) \sqrt{1+\alpha^2}\right) K(-\gamma) \right] , \\ 
\frac{\partial \chi_{1 1}}{\partial \omega^2}&=& -\int {d^3 k\over (2\pi)^3} {\xi_k^2\over 4 E_k^5} \non\\&=&   
\frac{ m^{3/2}}{24 \pi^2 |\mu|^{3/2}}  \frac{1}{\alpha^2+\alpha^4}\sqrt{-\text{sign}(\mu) + \sqrt{1+\alpha^2}} 
\left[  (-2-3 \alpha^2) E(-\gamma) \right. \nonumber \\ 
 && \left. + \left(3 + 3 \alpha^2 +\text{sign}(\mu) \sqrt{1+\alpha^2}\right) K(-\gamma) \right] , \\ 
 \frac{\partial \chi_{1 1}}{\partial q^2}&=& \int {d^3 k\over (2\pi)^3}\left( {\xi_k (\xi_k^2-2\D_0^2)\over 8 m E_k^5}+{5\D_0^2 \xi_k^2 k^2\over 24 m^2 E_k^7} \right)\non\\&=&   
\frac{ m^{1/2}}{72 \pi^2 |\mu|^{1/2}}  \frac{1}{\alpha^2+\alpha^4}\sqrt{-\text{sign}(\mu) + \sqrt{1+\alpha^2}} 
\left[  (4+ \alpha^2) \text{sign}(\mu)  E(-\gamma) \right. \nonumber \\ 
 && \left. +  \left(6\, \text{sign}(\mu) (1+\alpha^2) + (10+7 \alpha^2) \sqrt{1+\alpha^2}\right) K(-\gamma) \right] , 
 \end{eqnarray}
 \begin{eqnarray}
\frac{\partial \chi_\text{L}}{\partial q^2}  &=& -\int {d^3 k\over (2\pi)^3} {\D_0^2 k^4\over 20 m^4 E_k^5}\non\\&=&   \frac{ |\mu|^{1/2}}{30 \pi^2 m^{1/2}}  \frac{1}{\alpha^2}\sqrt{-\text{sign}(\mu) + \sqrt{1+\alpha^2}} 
\left[  (-4-3 \alpha^2) E(-\gamma) \right. \nonumber \\
 && \left. + \left(3 + 3 \alpha^2 - \text{sign}(\mu) \sqrt{1+\alpha^2}\right) K(-\gamma) \right] , \\
 \frac{\partial \chi_\text{L}}{\partial \omega^2} &=& 0, \\
 \label{eq:jnom}
 \frac{\partial \chi_{j n}}{\partial \omega}&=& -\int {d^3 k\over (2\pi)^3} {\D_0^2 k^2 \over 12 m^2 E_k^5}\non\\&=&  
\frac{ m^{1/2}}{36 \pi^2 |\mu|^{1/2}}  \frac{1}{\alpha^2}\sqrt{-\text{sign}(\mu) + \sqrt{1+\alpha^2}} 
\left[  -4\, \text{sign}(\mu) E(-\gamma) \right. \nonumber \\ 
 && \left.  +\left(3\, \text{sign}(\mu) -  \sqrt{1+\alpha^2}\right) K(-\gamma) \right] , \\ 
  \label{eq:j1om}
 \frac{\partial \chi_{j 1}}{\partial \omega}&=&-\int {d^3 k\over (2\pi)^3} {\D_0 \xi_k k^2\over 12 m^2 E_k^5}\non\\ &=&  
\frac{ m^{1/2} \Delta_0}{12 \pi^2 |\mu|^{3/2}}  \frac{1}{\alpha^2}\sqrt{-\text{sign}(\mu) + \sqrt{1+\alpha^2}} 
\left[ - E(-\gamma) + K(-\gamma) \right] .   
\end{eqnarray}
The derivatives of the transverse current response are in the same approximation
\begin{eqnarray}
\frac{\partial \chi_\text{T}}{\partial q^2}  &=& \frac{1}{3}\frac{\partial \chi_\text{L}}{\partial q^2}  , \\ 
\frac{\partial \chi_\text{T}}{\partial \omega^2}  &=& 0. 
\end{eqnarray}


\section{\label{sec:3point} One-loop three-point functions and the ratio $g_3/g_2$.}
In this Appendix we show how the  computation of a pair of  three-point functions in the one-loop approximation leads to the result  $g_3(X)/g_2(X)=2$. In order to identify the coefficients of $(\nabla^2 \theta)^2$ and $(\partial_i \partial_j \theta)^2$, 
we first write down all the third-order terms  proportional to $Y\theta \theta$ in the effective Lagrangian~(\ref{eq:lag6}) 
\begin{equation}
\mathcal{L}^{(3)} = -\frac{1}{m} g_5(\mu)  \bm{\nabla}\theta \cdot \bm{\nabla} Y \nabla^2 \theta  
-g_2'(\mu) Y (\nabla^2 \theta)^2 - g_3'(\mu) Y (\partial_i \partial_j \theta)^2, 
\end{equation} 
where $Y =  \mu-X$. This cubic Lagrangian gives rise to a  coupling  $Y \theta \theta$ with  vertex proportional to 
\begin{equation}\label{eq:vertex}
\frac{1}{m} g_5(\mu) \left[ \bm{q}_1 \cdot \bm{q}_2 (q_1^2+q_2^2) + 2 q_1^2 q_2^2 \right]
-2g_2'(\mu) q_1^2 q_2^2 -  2g_3'(\mu) (\bm{q}_1 \cdot \bm{q}_2)^2, 
\end{equation}
where $\bm{q}_1$ and $\bm{q}_2$ are the momenta of the pair of Goldstone fields.  
Note that the coefficient of the first term is known in terms of  two-point functions 
\beq\label{eq:g5}
g_5 = b_6-b_3=- \frac{\partial \chi_{j n}}{\partial \omega} + \frac{ g \chi_{n 1}}{2 + g \chi_{1 1}} \frac{\partial \chi_{j 1}}{\partial \omega}\;.
\eeq

Next, we compute the Fourier transforms of the one-loop three-point functions that  induce the effective vertex~(\ref{eq:vertex}). These are given by
$\langle T  \left\{ \Psi^\dagger(x) \tau^{1,3} \Psi(x) \bm{\nabla} \cdot \bm{j}^\text{p} (y) 
\bm{\nabla} \cdot \bm{j}^\text{p} (z) \right\} \rangle$,  
 where  $\bm{j}^\text{p}$  is the paramagnetic fermionic  current that couples  to $\bm{\nabla} \theta$ in Eq.~(\ref{eq:LI}).   
   The two triangle diagrams that contribute read 
\begin{eqnarray}
\Gamma_{1,3}^{(3)}(Q_1, Q_2)&=&\int_{K} \text{tr} \left[ \mathcal{G}(K) \mathcal{G}(K+Q_1)\mathcal{G}(K+Q_1+Q_2)   \tau^{1,3} \right] 
\nonumber \\ &\times&\left(  \bm{k} \cdot \bm{q}_1 + \frac{q_1^2}{2} \right)\left(   \bm{k} \cdot \bm{q}_2+ \bm{q}_1 \cdot \bm{q}_2+\frac{q_2^2}{2} \right)\frac{1}{m^2}  
+ \left( Q_1 \leftrightarrow  Q_2 \right). 
\end{eqnarray}
For our purposes, it is sufficient to obtain the $O(q^4)$ contribution to $\Gamma_{1,3}^{(3)}$ at  zero frequencies.  
A lengthly computation yields
\begin{eqnarray}
\Gamma_3^{(3)} &=& \int \frac{d^3 k}{(2 \pi)^3} {\D_0^2 k^2 \over 12 m^3 E_k^5}  
\left[   \bm{q}_1 \cdot \bm{q}_2 (q_1^2+q_2^2) + 2 q_1^2 q_2^2 \right] \nonumber \\ 
&&+\int \frac{d^3 k}{(2 \pi)^3}  \frac{\Delta_0^2 k^2 (  k^2 \xi_k- 2m E_k^2)}{12\, m^4 E_k^7} q_1^2 q_2^2  \nonumber \\ 
&&+\int \frac{d^3 k}{(2 \pi)^3} \frac{\Delta_0^2 k^2 (  k^2 \xi_k- 2m E_k^2)}{6\, m^4 E_k^7}   (\bm{q}_1 \cdot \bm{q}_2)^2 +  O(q^5), \\
\Gamma_1^{(3)} &=& \int {d^3 k\over (2\pi)^3} {\D_0 \xi_k k^2\over 12 m^2 E_k^5}  
\left[   \bm{q}_1 \cdot \bm{q}_2 (q_1^2+q_2^2) + 2 q_1^2 q_2^2 \right]  \nonumber \\ 
&&+\int \frac{d^3 k}{(2 \pi)^3} \frac{\Delta_0 k^2\left[ k^2 (6\,\xi_k^2+\Delta_0^2) -10 m \xi_k E_k^2 \right]}{60\, m^4 E_k^7}  q_1^2 q_2^2   \nonumber \\ 
&&+\int \frac{d^3 k}{(2 \pi)^3}   \frac{\Delta_0 k^2\left[ k^2 (6\,\xi_k^2+\Delta_0^2) -10 m \xi_k E_k^2 \right]}{30\, m^4 E_k^7} (\bm{q}_1 \cdot \bm{q}_2)^2 
+ O(q^5). 
\end{eqnarray}

We observe that the coefficients of $ (q_1^2+q_2^2) + 2 q_1^2 q_2^2$ in $\Gamma_3^{(3)} $ and $\Gamma_3^{(1)} $ are given respectively by $ -m^{-1}\partial \chi_{j n}/\partial \omega $ and $ -m^{-1}\partial \chi_{j 1}/\partial \omega $ (see Eqs.~(\ref{eq:jnom}) and (\ref{eq:j1om})).
This agrees with Eq.~(\ref{eq:g5}), after the replacement $\sigma \rightarrow g \chi_{11}Y/(2 + g \chi_{11}) $  ---from the gap equation--- is made in the term of the effective action proportional to  $\Gamma_1^{(3)} \sigma \theta \theta$. 
Simple inspection of the coefficients of  $(\bm{q}_1 \cdot \bm{q}_2)^2$ and $q_1^2 q_2^2$ shows that $g_3'(\mu)/g_2'(\mu) =2$, and therefore $g_3(X)/g_2(X)= 2$, after dropping an irrelevant   integration constant.

The result $g_3(X)/g_2(X)= 2$  is strikingly simple, and one might  wonder whether  this is just a peculiarity of the one-loop aproximation used here. But, unlike the ratio $f_3/f_2=-3$ found in Section~\ref{sec:higher}, the value  $g_3(X)/g_2(X)= 2$ does not seem to be  the consequence  of any obvious symmetry.


\end{document}